\documentclass[pdflatex,sn-nature]{sn-jnl}

\usepackage{graphicx}%
\usepackage{multirow}%
\usepackage{amsmath,amssymb,amsfonts}%
\usepackage{amsthm}%
\usepackage{mathrsfs}%
\usepackage[title]{appendix}%
\usepackage{xcolor}%
\usepackage{textcomp}%
\usepackage{manyfoot}%
\usepackage{booktabs}%
\usepackage{algorithm}%
\usepackage{algorithmicx}%
\usepackage{algpseudocode}%
\usepackage{listings}%
\usepackage{comment}
\newcommand{\Add}[1]{\textcolor{black}{#1}}

\newcommand{\aj}{Astron. J.}   
\newcommand{\apj}{Astrophys. J.}   
\newcommand{\apjl}{Astrophys. J. Lett.}   
\newcommand{\apjs}{Astrophys. J. Suppl. Ser.}   
\newcommand{\aap}{Astron. Astrophys.}   
\newcommand{\mnras}{Mon. Not. R. Astron. Soc.}   
\newcommand{\pasj}{Publ. Astron. Soc. Jpn}   
\newcommand{\pasp}{Publ. Astron. Soc. Pac.}   
\newcommand{\solphys}{Sol. Phys.}   
\newcommand{\ssr}{Space Sci. Rev.}   

\theoremstyle{thmstyleone}%
\theoremstyle{thmstyletwo}%

\theoremstyle{thmstylethree}%

\begin{document}

\title[Article Title]{Discovery of multi-temperature coronal mass ejection signatures from a young solar analogue}


\author*[1,2,3,4,5]{\fnm{Kosuke} \sur{Namekata}}\email{namekata@kusastro.kyoto-u.ac.jp}

\author[6,7]{\fnm{Kevin} \sur{France}}
\author[8,9]{\fnm{Jongchul} \sur{Chae}}
\author[1,10]{\fnm{Vladimir S.} \sur{Airapetian}}
\author[6,7,11]{\fnm{Adam} \sur{Kowalski}}
\author[6,11]{\fnm{Yuta} \sur{Notsu}}
\author[1,12]{\fnm{Peter R.} \sur{Young}}
\author[13]{\fnm{Satoshi} \sur{Honda}}
\author[8]{\fnm{Soosang} \sur{Kang}}
\author[8,9]{\fnm{Juhyung} \sur{Kang}}
\author[8]{\fnm{Kyeore} \sur{Lee}}
\author[14]{\fnm{Hiroyuki} \sur{Maehara}}
\author[8,9]{\fnm{Kyoung-Sun} \sur{Lee}}
\author[6,11]{\fnm{Cole} \sur{Tamburri}}
\author[13,15]{\fnm{Tomohito} \sur{Ohshima}}
\author[13,16]{\fnm{Masaki} \sur{Takayama}}
\author[17,18]{\fnm{Kazunari} \sur{Shibata}}

\affil[1]{\orgdiv{Heliophysics Science Division}, \orgname{NASA Goddard Space Flight Center}, \orgaddress{\street{8800 Greenbelt Road}, \city{Greenbelt}, \postcode{20771}, \state{MD}, \country{USA}}}

\affil[2]{\orgdiv{Department of Physics}, \orgname{The Catholic University of America}, \orgaddress{\street{620 Michigan Avenue, N.E.}, \city{Washington}, \postcode{20064}, \state{DC}, \country{USA}}}

\affil[3]{\orgdiv{The Hakubi Center for Advanced Research}, \orgname{Kyoto University}, \orgaddress{\street{Yoshida-Honmachi, Sakyo-ku}, \city{Kyoto}, \postcode{606-8501}, \state{Kyoto}, \country{Japan}}}

\affil[4]{\orgdiv{Department of Physics}, \orgname{Kyoto University}, \orgaddress{\street{Kitashirakawa-Oiwake-cho, Sakyo-ku}, \city{Kyoto}, \postcode{606-8502}, \state{Kyoto}, \country{Japan}}}

\affil[5]{\orgdiv{Division of Science}, \orgname{National Astronomical Observatory of Japan, NINS}, \orgaddress{\street{Osawa, Mitaka}, \city{Tokyo}, \postcode{181-8588}, \state{Tokyo}, \country{Japan}}}

\affil[6]{\orgdiv{Laboratory for Atmospheric and Space Physics}, \orgname{University of Colorado Boulder}, \orgaddress{\street{3665 Discovery Drive}, \city{Boulder}, \postcode{80303}, \state{CO}, \country{USA}}}

\affil[7]{\orgdiv{Department of Astrophysical and Planetary Sciences}, \orgname{University of Colorado Boulder}, \orgaddress{\city{Boulder}, \postcode{80309}, \state{CO}, \country{USA}}}

\affil[8]{\orgdiv{Astronomy Program, Department of Physics and Astronomy}, \orgname{Seoul National University}, \orgaddress{\street{Gwanak-gu}, \city{Seoul}, \postcode{08826}, \country{Korea}}}

\affil[9]{\orgdiv{Astronomy Research Center}, \orgname{Seoul National University}, \orgaddress{\street{Gwanak-gu}, \city{Seoul}, \postcode{08826}, \country{Korea}}}

\affil[10]{\orgdiv{Sellers Exoplanetary Environments Collaboration}, \orgname{NASA Goddard Space Flight Center}, \orgaddress{\street{8800 Greenbelt Road}, \city{Greenbelt}, \postcode{20771}, \state{MD}, \country{USA}}}

\affil[11]{\orgdiv{National Solar Observatory}, \orgaddress{\street{3665 Discovery Drive}, \city{Boulder}, \postcode{80303}, \state{CO}, \country{USA}}}

\affil[12]{\orgname{Northumbria University}, \orgaddress{\city{Newcastle upon Tyne}, \postcode{NE1 8ST}, \country{UK}}}

\affil[13]{\orgdiv{Nishi-Harima Astronomical Observatory, Center for Astronomy}, \orgname{University of Hyogo}, \orgaddress{\city{Sayo}, \state{Hyogo}, \postcode{679-5313}, \country{Japan}}}

\affil[14]{\orgdiv{Okayama Branch Office, Subaru Telescope}, \orgname{National Astronomical Observatory of Japan, NINS}, \orgaddress{\street{Kamogata, Asakuchi}, \city{Okayama}, \postcode{719-0232}, \country{Japan}}}

\affil[15]{\orgdiv{Solar Science Observatory}, \orgname{National Astronomical Observatory of Japan, NINS}, \orgaddress{\street{Osawa, Mitaka}, \city{Tokyo}, \postcode{181-8588}, \state{Tokyo}, \country{Japan}}}

\affil[16]{\orgdiv{Center for Student Success Research and Practice}, \orgname{University of Osaka}, \orgaddress{\street{Toyonaka, Osaka}, \postcode{560-0043}, \country{Japan}}}

\affil[17]{\orgdiv{Kwasan Observatory}, \orgname{Kyoto University}, \orgaddress{\city{Yamashina, Kyoto}, \postcode{607-8471}, \country{Japan}}}

\affil[18]{\orgdiv{School of Science and Engineering}, \orgname{Doshisha University}, \orgaddress{\city{Kyotanabe, Kyoto}, \postcode{610-0321}, \country{Japan}}}





\abstract{\normalsize
Coronal mass ejections (CMEs) on the early Sun may have profoundly influenced the planetary atmospheres of early Solar System planets.
Flaring young solar analogues serve as excellent proxies for probing the plasma environment of the young Sun, yet their CMEs remain poorly understood.
Here we report the detection of multi-wavelength Doppler shifts in Far-Ultraviolet (FUV) and optical lines during a flare on the young solar analog EK Draconis.
During and before a Carrington-class ($\sim$10$^{32}$ erg) flare, warm FUV lines ($\sim$10$^5$ K) exhibit blueshifted emission at 300-550 km s$^{-1}$, indicative of a warm eruption.
10 minutes later, the H$\alpha$ line shows slow (70 km s$^{-1}$), long-lasting ($\gtrsim$2 hrs) blueshifted absorptions, suggesting a cool ($\sim$10$^4$ K) filament eruption.
This provides evidence of multi-temperature and multi-component nature of a stellar CME.
If Carrington-class flares/CMEs occurred frequently on the young Sun, they may have cumulatively impacted the early Earth’s magnetosphere and atmosphere.
}

\maketitle



Solar filaments--cool, dense plasma (10$^{4-5}$ K \Add{and 10$^{9-11}$ cm$^{-3}$}) in the hot corona--play a key role in diagnosing coronal magnetic destabilization and driving space weather in the Solar System \citep{2014LRSP...11....1P}.
When destabilized in association with flares, they often erupt and evolve into coronal mass ejections (CMEs) \citep{2011LRSP....8....1C}, exhibiting complex multi-temperature, multi-density structures.
These eruptions can generate shock waves and accelerate energetic particles, significantly impacting planetary magnetospheres and unmagnetized atmospheres like that of Mars \citep{2015Sci...350.0210J}.
The discovery of numerous exoplanets around young G-, K-, and M-type stars highlights the potential role of stellar CMEs in shaping planetary atmospheres.
Young solar analogues, in particular, offer a unique window into both the evolution of Earth-like atmospheres and the early conditions that may have supported the origin of life.
These stars are far more active than the modern Sun \citep{2012Natur.485..478M}, making frequent flares and CMEs likely drivers of planetary evolution.
Their influence may extend beyond atmospheric loss \citep{2022MNRAS.509.5858H}, affecting the efficient formation of prebiotic chemistry, greenhouse gases \citep{2016NatGe...9..452A}, and amino \Add{acids} \citep{2023Life...13.1103K}.
Although theory and experiments suggest such effects, direct observational evidence for eruptive phenomena on young solar analogues remains scarce.

Various observational techniques have been employed to study stellar eruptive phenomena, but clear signatures are often obscured by the brightness of host stars and flares--leaving only ambiguous evidence, especially for solar analogues.
To date, Doppler shifts in optical lines have yielded only a few likely detections of stellar filament eruptions associated with giant superflares on a young solar analogue \citep{2022NatAs...6..241N,2024ApJ...961...23N}.
However, these detections have been limited to single-wavelength data at low temperatures typical of filaments or prominences ($\sim10^4$ K), leaving the overall multi-temperature structure largely unexplored, especially fast and hot plasma at greater altitudes.
As a result, CMEs on solar analogues remain poorly constrained in terms of planetary impact.
This challenge extends to M-dwarfs and close binaries as well.
While X-ray/UV/optical Doppler shift studies \citep{2019NatAs...3..742A,2024ApJ...961..189N} and other promising techniques, such as X-ray/UV dimming \citep{2021NatAs...5..697V,2025arXiv250519228V,2025ApJ...988..167M}, have revealed more insights in these cool stars, simultaneous multi-wavelength observations remain unsuccessful.






Among various methods, Doppler shifts are, if interpreted properly, an extremely potent tool for constraining explosive mass motions and/or ejections, particularly their multi-thermal nature.
On the Sun, Doppler shifts of eruptive events are commonly observed in ultraviolet and optical/infrared wavelengths, tracing warm ($\sim 10^5$ K) and cool ($\sim 10^4$ K) plasma, respectively \citep{2015ApJ...806....9K,2022NatAs...6..241N}.
While hot coronal plasma emitting in extreme-ultraviolet (EUV) and X-rays is ejected with CME flux ropes, other flare-related upflows, known as chromospheric evaporation, also produce X-ray Doppler shifts at $\gtrsim 10^6$ K \citep{2009ApJ...699..968M}, introducing ambiguity in Sun-as-a-star spectral interpretations.
Simultaneous ultraviolet and optical/infrared spectroscopy is therefore ideal for probing the multi-temperature structure of stellar eruptive events.
To this end, we conducted multi-wavelength monitoring observations of the nearby young solar analogue EK Draconis (EK Dra) using the Hubble Space Telescope (HST), the Transiting Exoplanet Survey Satellite (TESS), and three ground-based telescopes in Japan and South Korea.
The campaign spanned four consecutive nights from March 29 to April 1, 2024.
EK Dra (HD 129333, G1.5V) is a young solar analogue with an estimated age of 50--125 Myr. Its effective temperature (5560--5700 K), radius (0.94 $R_\odot$), and mass (0.95 $M_\odot$) make it one of the best proxies for the infant Sun \citep{2022NatAs...6..241N,2024ApJ...961...23N}.

\section*{Results}

\begin{figure}[!hb] 
\begin{center}
\includegraphics[width=1.0\linewidth,clip]{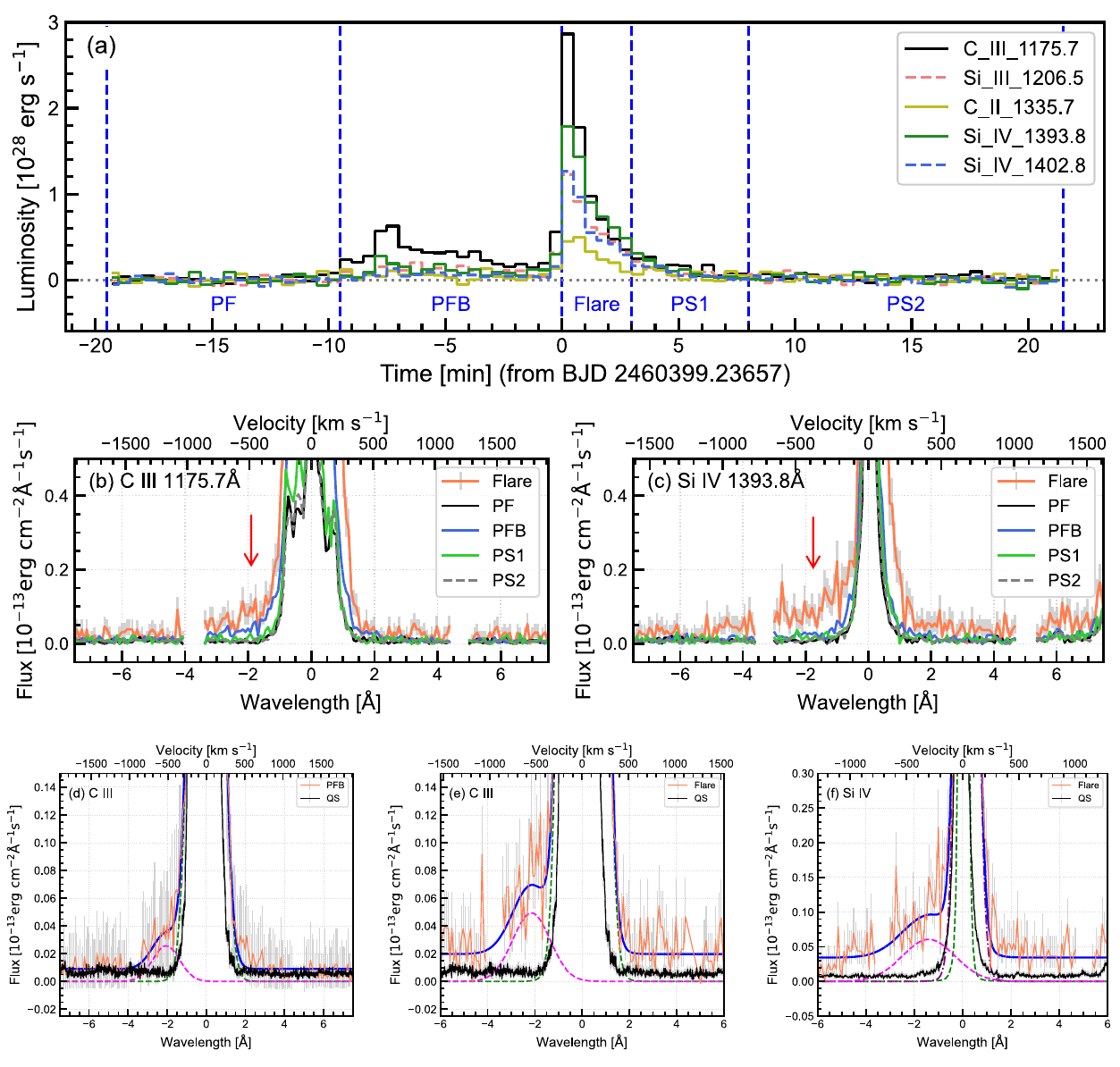}
\caption{The FUV spectral lines showing blueshifted emission components.  
(a) Light curves of pre-flare subtracted luminosity of the top five strongest FUV spectral lines in the third HST's orbit during the March 29, 2024. 
We divided the light curve into five phases--pre-flare (PF), pre-flare brightening (PFB), flare, post-flare 1 (PS1), and post-flare 2 (PS2)--to extract phase-specific spectra.  
(b, c) Temporal evolution of the C III 1175.7 Å (log$T$=4.85) and Si IV 1393.8 Å (log$T$=4.85) spectra, which exhibited prominent blueshifted components. As a reference, error bars are shown for the flare-phase data. To enhance the signal-to-noise ratio (S/N), the spectra were binned over eight wavelength pixels (i.e., 0.080 {\AA} = 20 km s$^{-1}$ for C III and 0.080 {\AA} = 17 km s$^{-1}$ for Si IV).  
(d) Gaussian fitting of the C III spectral line during the PFB phase. The flare component at the line center is fitted with a single Gaussian (green dashed line), while the residual blueshifted component is fitted separately with another Gaussian (magenta dashed line).
The blue line represents the total line profile, obtained by combining the two Gaussian components.
(e) Gaussian fitting of the C III spectral line during the flare phase. 
The blueshifted component is a factor of 0.024 smaller than the rest line component.
(f) Gaussian fitting of the Si IV spectral line. To reproduce the complex redshifted profile, a double-Gaussian fit was applied to the central component \Add{as indicated by the green dashed and purple dash-dotted lines}.
The blueshifted component is a factor of 0.016 smaller than the rest line component.
}
\label{fig:fuv}
\end{center}
\end{figure}

On March 29, 2024, we detected a large flare in the far-ultraviolet (FUV) from the young solar analogue EK Dra.
Figure \ref{fig:fuv}(a) shows light curves of five strong FUV lines observed by HST: C III 1175.7 \AA, Si III 1206.5 \AA, C II 1335.7 \AA, and Si IV 1393.8 \AA\ and 1402.8 \AA.
The event included a pre-flare brightening (``PFB'') and a major flare.
Most lines show overall redshifts \Add{(cf. Extended Data Figures 1 and 2)}, typical of chromospheric downflows.
In contrast, C III ($\log T = 4.85$) and Si IV ($\log T = 4.85$) display extended blueshifted wings during the flare (Figure \ref{fig:fuv}b,c), with C III showing a blueshift during the PFB phase.
Other FUV lines suffer from line blending or poor data quality, but possible blueshift signatures are also seen in the blue wing of Ly$\alpha$, the C II 1334.5/1335.7 \AA\ doublet, and Si IV 1402.8 \AA\ (Extended Data Figure 2).
In addition, one minor flare on another night also showed FUV blueshifts, while another did not (Extended Data Figure 3), suggesting such features may be frequent.
We fitted both central and blueshifted spectra with multiple Gaussian components (see Methods for details).
As shown in Figure~\ref{fig:fuv}(d--f), the blueshifted Si IV component during the flare shows a \Add{central} velocity of $-300 \pm 60$ km s$^{-1}$, with a maximum blueshift of $-470 \pm 70$ km s$^{-1}$.
C III emission consists of five blended lines; we estimated velocities from the brightest (1175.7 \AA) and bluest (1174.9 \AA) components.
During the flare, their central blueshifted velocities are $-550 \pm 50$ km s$^{-1}$ and $-350 \pm 50$ km s$^{-1}$, with maxima of $-690 \pm 70$ km s$^{-1}$ and $-500 \pm 70$ km s$^{-1}$, respectively.
Blueshifted velocities in the PFB phase are comparable to those in the flare phase.
The velocity from the bluest C III line \Add{is consistent with} that of Si IV.

As shown in Figure \ref{fig:ha}, about ten minutes after the FUV flare, a blueshifted absorption with velocities of $-60$ to $-70$ km s$^{-1}$ appeared in the H$\alpha$ line at 6562.8 Å and persisted for about two hours until the end of observations. Although the exact onset is uncertain, the blueshift was absent during the FUV flare itself. This feature was confirmed by two high-resolution spectrographs: BOES on the 1.8 m BOAO telescope in South Korea and MALLS on the 2.0 m Nayuta telescope in Japan (see Extended Data Fig. 4).
Notably, the H$\alpha$ line also showed a separate, long-duration flare, lasting from $-200$ to $-50$ minutes before the FUV flare, without corresponding FUV emission. This earlier flare had nearly decayed by the time of the FUV event. Its H$\alpha$ profile exhibits broadened, redshifted emission, typical of solar and stellar flares \citep{2022ApJ...933..209N}, and is attributed to chromospheric condensation \Add{(see Methods Section ``Emission Mechanism of the Long-Duration H$\alpha$ Flare" for detailed interpretations)}.
The physical connection between this precursor H$\alpha$ flare and the FUV/H$\alpha$ blueshifts remains unclear. 
However, the temporal proximity between the FUV and H$\alpha$ blueshifts suggests a possible causal relationship. 
This observation constitutes \Add{a} multi-wavelength detection of blueshifts in both the FUV and optical regimes.

\section*{Discussion}


From the flaring FUV spectrum (Extended Data Fig. 5), the bolometric white-light flare energy is estimated at $2.3 \pm 1.1 \times 10^{32}$ erg, corresponding to an $\sim$X23-class flare in the solar GOES classification (see Methods).
This energy scale is consistent with the absence of significant flaring signatures in TESS optical photometry and H$\alpha$, which have detection thresholds of $\sim 10^{33}$ erg \citep{2022NatAs...6..241N,2024ApJ...961...23N} (Extended Data Figure 6).
A useful comparison is the 1859 ``Carrington” flare--the largest solar flare on record--which is $\sim$X45 class \citep{2022LRSP...19....2C}.
While this event is extreme for the Sun, EK Dra is a highly active young star, producing superflares of $\gtrsim 10^{33}$ to $\sim 5 \times 10^{34}$ erg roughly every two days \cite{2022ApJ...926L...5N}.
Thus, the flare we detected is relatively ``minor” for EK Dra, but remains valuable as a proxy for extreme solar space weather.

We report, in a young solar analogue, the detection of FUV blueshifted components in multiple emission lines during flares. 
These FUV blueshifts suggest the possible occurrence of coronal mass ejections (CMEs), by analogy with solar events.
On the Sun, filament eruptions and jets often appear as blueshifted FUV emissions, including in Si \textsc{iv} \citep{2015ApJ...806....9K,2021A&A...647A.113Z,2022A&A...660A..45M}. Recent Sun-as-a-star observations with the Extreme Ultraviolet Variability Experiment onboard the Solar Dynamics Observatory show that CMEs can be traced in EUV/FUV lines formed at $\log T = 4.2$--$5.8$, particularly during the flare's early phase \citep{2022ApJ...931...76X,2024ApJ...964...75O}. 
This range matches the formation temperatures of the observed C \textsc{iii} and Si \textsc{iv} lines ($\log T = 4.85$).
The short duration of the FUV blueshifts may be explained by the rapid cooling or expansion of the $\sim$$10^5$ K plasma.

We believe alternative explanations for the FUV blueshifts to be unlikely.
First, no blended lines are present in the quiescent spectra of the C~\textsc{iii} and Si~\textsc{iv} lines, nor have flare-sensitive features been reported at these wavelengths.
Second, while small-scale solar events such as ``UV bursts" can show strong FUV blueshifts \citep{2018SSRv..214..120Y}, their persistent nature in active regions makes them an unlikely cause.
Third, chromospheric evaporation during solar flares can drive upflows of hot plasma (10$^6$--10$^7$ K), but these produce blueshifts \Add{mostly} in coronal lines formed at $\gtrsim 10^6$ K \citep{2009ApJ...699..968M}.
\Add{Occasionally, blueshifts in flare ribbons are observed in $\sim2\times10^5$ K EUV O V lines \citep{2010ApJ...719..655V}, but they are rare, short lived, and localized. They are even rarer in the FUV and would be too faint in Sun-as-a-star or stellar spectra to account for the observed emission measure (see Methods Section ``Plasma Mass Estimation"). 
These data suggest that the blueshifts observed in FUV emission lines forming at $\sim$10$^5$ K are unlikely from the flare-driven evaporation process.} 
As a comparison, Extended Data Fig. 7 shows Si \textsc{iv} and C \textsc{ii} spectra for an X9.0-class solar flare with comparable energy. This flare near disk center does not show eruptive filaments and exhibits no extended blueshifted wings, confirming that chromospheric evaporation minimally affects these warm lines.
Although numerical simulations of FUV line dynamics remain limited \citep{2005ApJ...630..573A}, current solar observations support the interpretation that the FUV blueshifts in EK Dra are most consistent with flare-associated CMEs.



The plasma dynamics observed in H$\alpha$ are best explained by a filament eruption, analogous to those seen on the Sun \citep{2022NatAs...6..241N,2022ApJ...939...98O,2025arXiv250707967D}. As the filament rises against the stellar disk, it produces a blueshifted absorption profile. The observed velocity of $-60$ to $-70$ km s$^{-1}$ is consistent with solar filament eruptions \citep{2022ApJ...939...98O} and significantly exceeds the stellar rotational velocity of $16.4 \pm 0.1$ km s$^{-1}$ \citep{2017MNRAS.465.2076W}.
On the Sun, post-flare loop downflows may show blueshifts near the limb at certain viewing angles \citep{2024ApJ...974L..13O}, but redshifted absorption generally dominates, making \Add{the post-flare loop downflow} explanation unlikely. 
\Add{Numerical simulations by Ref. \cite{2025A&A...694A.315Y} show blueshifted absorption from cool upward flows in flare ribbons, but this mechanism is unlikely to explain the stellar data, as it has not been observed in spatially resolved solar observations and cannot account for the long-lasting post-flare behavior.}  
Thus, a filament eruption remains the most plausible interpretation.
Two novel issues remain: First is the possible connection with a long-duration H$\alpha$ flare that occurred 200--50 minutes prior. However, the flare had nearly decayed by the time of the blueshift, and filament-related blueshifts typically appear early in the flare \citep{2022NatAs...6..241N,2024ApJ...961...23N,2022ApJ...939...98O}, making a causal link unlikely.
Second, the H$\alpha$ blueshift persisted with a nearly constant velocity for about two hours, much longer than previous events ($\lesssim$20 min) on EK Dra \citep{2022NatAs...6..241N,2024ApJ...961...23N}. 
Although long-duration, slowly rising filaments have been seen in solar radio data \citep{2003ApJ...586..562G}, their mechanism remains unclear and such behavior is not well supported by \Add{both pseudo \citep{2022ApJ...939...98O} and true \citep{2024A&A...682A..46P}} Sun-as-a-star H$\alpha$ observations.
This prolonged blueshift may be explained by continuous external forcing acting on the filament, or by the preferential fading of faster components, leaving behind only the slower and denser material.
\Add{At least, the long-lasting nature is unlikely to be explained by confined eruptions \citep{2018ApJ...862...93A}, but is more consistent with CME events.}
While the detailed dynamics remain uncertain, the near-simultaneous detection of a H$\alpha$ filament eruption and FUV CME signatures suggests a physical association rather than a chance coincidence.




A key question is how the cool and warm plasma components are connected.
The timing and velocity evolution of the blueshifts in the FUV and H$\alpha$ lines are difficult to reconcile with a single plasma evolution scenario, such as a warm-to-cool eruptive transition.
First, the H$\alpha$ blueshift appears significantly later than the FUV flare, unlike solar flares where UV and H$\alpha$ emissions typically rise together \citep{2024ApJ...974L..13O}.
While this challenges the interpretation of a single growing filament, it does not entirely rule it out, as the H$\alpha$ component may have existed earlier but remained hidden in the line core due to low velocity.
Second, as shown in Figure \ref{fig:vel}, the velocity transitions from warm plasma at --300 to --550 km s$^{-1}$ to cooler plasma at --60 to --70 km s$^{-1}$ within 10 minutes.
Assuming a single plasma decelerating, this implies a deceleration of $>0.60$ km s$^{-2}$, far exceeding the surface gravity of $0.30 \pm 0.05$ km s$^{-2}$.
This large discrepancy with solar expectations strongly suggests that the FUV and H$\alpha$ blueshifts originate from distinct plasma components.



Based on the above discussion, we conclude that the blueshift phenomena observed in the FUV and H$\alpha$ lines are likely related, but originate either (1) from different atmospheric layers of a single event, or (2) from distinct yet connected eruptive events.
Figure \ref{fig:vel}(b) illustrates these two scenarios schematically.
In both cases, our results offer evidence for multi-temperature structures in stellar CME-related phenomena, involving fast, warm plasma and slow, cool plasma.
In scenario (1), this represents a detection of a multi-layer CME in a stellar event, analogous to solar eruptions, \Add{which consists of a low-temperature filament core and surrounding, upper-layer hot plasma}.
In scenario (2), the nearly simultaneous timing suggests that the FUV eruption may have triggered the H$\alpha$ eruption, resembling ``sympathetic” eruptions seen on the Sun \citep{2001ApJ...559.1171W}. This potentially represents a case of such sympathetic eruptions identified in stellar observations.
Importantly, scenario (2) also implies that some stellar eruptions may only be observable at certain wavelengths, i.e. optical or FUV in this case, indicating that CME occurrence rates based on single-wavelength data may substantially underestimate their true frequency on young solar analogues.
\Add{Magnetically active stars like EK Dra are tightly covered by large active regions \citep{1995A&A...301..201G}, so sympathetic eruptions may be much more common.}
\Add{A third possibility that the FUV and H$\alpha$ blueshifts occur independently cannot be completely ruled out, but the probability of two such events (at a rate of $\sim$5 per day) occurring within 10 minutes is 3.4\% with Poisson distribution, making this scenario statistically unlikely.}

The multi-temperature detection of blueshifted signatures in both FUV and H$\alpha$ allows independent estimates of the mass and kinetic energy of each component, shedding light on the connection between plasmas at different temperatures.
The estimated mass of the fast, warm plasma is $(4.0$–$7.6) \times 10^{16}$ g, while that of the slow, cool plasma is $(0.93$–$31) \times 10^{16}$ g (see Methods), suggesting that both components carry comparable mass from the star and may be physically linked.
In contrast, the warm plasma \Add{likely carries} more kinetic energy, $(1.1$–$4.9) \times 10^{31}$ erg, than the cool plasma, $(0.19$–$9.0) \times 10^{30}$ erg.
In scenario (1), this energy difference is consistent with the solar self-similar CME framework, where higher-temperature plasma at greater heights is ejected more rapidly.
In scenario (2), it is plausible that an initial, more energetic FUV eruption triggered the subsequent, less energetic H$\alpha$ eruption.

To place the observed eruptions in a solar context, we compared the mass and kinetic energy of solar and stellar eruptions as a function of flare energy (see Extended Data Figure 8). The estimated masses align well with solar scaling relations, suggesting similar underlying physical processes.
In contrast, the kinetic energy of the cool plasma is about two orders of magnitude lower than predicted by empirical solar scaling laws \citep{2013ApJ...764..170D}, while that of the warm plasma is more consistent with the expected relation.
Two explanations have been proposed for the low kinetic energy of H$\alpha$ filaments: (1) the cool component represents the lower, slower part of the CME \citep{2022NatAs...6..241N}, and (2) the eruption is suppressed by the strong overlying coronal field in active stars \citep{2018ApJ...862...93A}.
Our findings support the former scenario, in which the warm, high-velocity plasma carries most of the kinetic energy, consistent with solar flare-CME scaling laws.


Our findings provide key insights into the energy and frequency of flares associated with CME-related phenomena. We observed such events during one flare of $\sim$10$^{32}$ erg (X10-class) and one of two flares of $\sim$10$^{31}$ erg (X-class) (Figure \ref{fig:fuv}, Extended Data Figure 3).
Previous H$\alpha$ observations suggested a high CME association rate for stellar superflares ($\gtrsim$10$^{33}$ erg), with four out of 15 events showing CME signatures, noting that a large fraction of ejections directed perpendicular to the line of sight would likely be missed \citep{2024ApJ...961...23N,namekata2025apj}.
Our results indicate that CME-related eruptions are relatively common even at lower energies above $\sim$10$^{31}$ erg.
Moreover, since some eruptions may only be detectable in specific wavelengths, the true occurrence rate is likely even higher.
This has important implications: such eruptions occur even in Carrington-class flares, which are 2--3 orders of magnitude weaker than the largest flares observed on young solar analogues.
Although simulations suggest that strong dipolar fields can suppress smaller eruptions \citep{2018ApJ...862...93A}, young solar analogues such as EK Dra exhibit predominantly quadrupolar fields, resulting in a mix of open and closed regions \citep{2024ApJ...976..255N}.
In such configurations, magnetic suppression is likely less effective, which is consistent with a numerical simulation by Ref.  \cite{2024MNRAS.533.1156S}, who show that suppression is significantly weaker in open field regions.
Our results support the idea that eruptions originate in these open regions and may occur frequently.

If such minor, Carrington-class CMEs were common on the young Sun, their cumulative impact on planetary atmospheres may warrant re-evaluation.
Optical and X-ray data suggest that Carrington-class flares occurred at a rate of $\sim$5--10 per day on this star \citep{2022ApJ...926L...5N}.
If each flare was accompanied by a CME, the young Earth may have experienced magnetic storms $\sim$0.5--1 times per day.
Fast CMEs are expected to drive strong shocks that accelerate particles to GeV energies via diffusive shock acceleration \citep{2022SciA....8I9743H}.
Given CME transit times of $\sim$2 days, the early Earth may have been under near-constant shock exposure.
The enhanced dynamic pressure could have compressed the magnetosphere to $\sim$2 $R_{\rm E}$, expanding the polar cap to $\sim$30° latitude and allowing energetic particles to precipitate efficiently \citep{2016NatGe...9..452A,2020IJAsB..19..136A}.
These particles would ionize atmospheric N$_2$ and CO$_2$, generating free radicals, precursors to amino acids, carboxylic acids, and nitrous oxide \citep{2023Life...13.1103K}.
Such conditions could have fostered prebiotic chemistry, linking solar activity to the origin of life on Earth.
\Add{From another perspective, CME mass loss rates from Carrington-class events could reach values as high as $\sim$10$^{-13}$ $M_\odot$ yr$^{-1}$, potentially impacting the early Sun’s angular momentum and mass evolution \citep{2012ApJ...760....9A,2013ApJ...764..170D}.}

We detected a stellar flare accompanied by CME-related phenomena in both far-ultraviolet and optical wavelengths. This provides evidence of either multi-temperature, multi-layer eruptions or sympathetic eruptions.
Our findings combined with the earlier work of \cite{2022NatAs...6..241N,2024ApJ...961...23N} suggest that powerful CMEs occur frequently on young solar analogues, implying a greater cumulative impact on the atmospheres of early Earth and exoplanets orbiting young solar analogues.
Furthermore, this study highlights the importance of UV time-domain astronomy and the need for extended monitoring with HST or future missions like JAXA's LAPYUTA, a far-UV observatory designed to detect such eruptions with high spectral resolution \citep{2024SPIE13093E..0IT}, \Add{to further characterize their statistical properties such as frequency}.
Investing in UV observations of solar-analogue host stars is essential groundwork for upcoming missions like NASA’s Habitable Worlds Observatory, which seeks to characterize the atmospheres of potentially habitable exoplanets around them.


\clearpage


\clearpage

\backmatter





\section*{Methods}

\subsection*{FUV Spectroscopic Data Analysis}

The Cosmic Origins Spectrograph (COS) onboard HST was utilized for our observation campaign \citep[HST Proposal. Cycle 31, ID. \#17595;][]{2024hst..prop17595F}.
The FUV G130M grating was used with a central wavelength of 1291 {\AA}. 
The spectral coverage of G130M is about 1130--1430 {\AA}, and the spectral resolving power is 12,000--16,000 (\url{https://hst-docs.stsci.edu/cosihb/chapter-13-cos-reference-material/13-3-gratings/fuv-grating-g130m\#FUVGratingG130M-Note1}).
The instrument was operated at Lifetime Position 5.
The COS data were obtained from the Mikulski Archive for Space Telescopes (MAST). The photons are time-tagged, allowing for the extraction of spectra and light curves with arbitrary time cadences. We used the \texttt{splittag} function in the Python package \texttt{costools} to split the files named ``corrtag" into a time cadence (\(\Delta t_{\rm step}\)) of 30 s. 
Finally, we used \texttt{calcos} (version 3.4.8) to generate flux-calibrated spectra.
After the \texttt{calcos} calibration, the spectrum was corrected for the line-of-sight velocity to account for the stellar proper motion of -20.687 km s$^{-1}$.
We excluded bad pixels based on the ``DQ" flags in the \texttt{calcos} output. 

The ``EXPSTART'' in the \texttt{calcos} output's FITS header provides the exposure start time for each orbit, based on which we calculate the time of \(n\) as ``EXPSTART'' + \(n \times \Delta t_{\rm step} + 0.5 \Delta t_{\rm step}\).
To align with the TESS provided in BJD, the UTC timestamps from the HST data were converted to BJD using the coordinates of Greenwich and the target object.
When constructing the light curve in Figure \ref{fig:fuv}(a), we integrated the flux over \(\pm1.0\) {\AA} for C III to include all severely blended lines, \(\pm0.75\) {\AA} for Si III and Si IV, and \(\pm0.5\) {\AA} for C II.
After reviewing the Si IV light curves, we identified the ``FUV quiescent phases" in FUV as orbits 1 and 2 in visit 1, orbits 1 and 3 in visit 2, orbits 1, 2, and 3 in visit 3, and orbits 1 and 2 in visit 4, and made a template spectrum by averaging these orbits, as seen in Figure \ref{fig:fuv} (d,e,f).
We primarily use this high S/N FUV quiescent spectrum as a template for comparison with the flaring spectrum.

The \texttt{calcos} pipeline provides asymmetric error values for the upper (``ERROR") and lower (``ERROR\_LOWER") flux uncertainties based on the Poisson photon noise \citep{2021cos..rept....3J}.
When integrating the time-resolved spectra along the wavelength and time axes, we applied the same error propagation calculation as the \texttt{calcos} pipeline. Specifically, we summed the values of ``VARIANCE\_FLAT", ``VARIANCE\_COUNTS", and ``VARIANCE\_BKG", and determined the upper and lower error bars for the count values using \texttt{astropy.stats.poisson\_conf\_interval} with the condition \texttt{interval=`frequentist-confidence'}.
To sufficiently improve the signal-to-noise ratio, we applied temporal and wavelength binning. As a result, the upper and lower error bars were nearly symmetrical. Therefore, we used the upper error bar, which is larger than the lower, for spectral fitting and error range evaluation.

For the spectral line fitting, we followed the instruction in the STScI HST Notebook (see Code Availability) considering the line spreading function (LSF) of COS.
We use an LSF file (aa\_LSFTable\_G130M\_1291\_LP5\_cn.dat) based on the COS Lifetime Position (LP) of 5, grating of G130M, and central wavelength setting (cenwave) of 1291 {\AA} and a DISPTAB file (5992009ol\_disp.fits) based on the fits header.
We use \texttt{convolve\_lsf} function in the LSF Notebook (see Code Availability) with input of the above parameters and files.
\Add{We fitted the C III line center component with a single Gaussian and the Si IV line with a double Gaussian. For C III, which is a complex line, we isolated the blue-wing residual by fitting only the line wing of the central component with a single Gaussian. Si IV, although sufficiently isolated, is known to be best represented by a narrow and a broad Gaussian component (as shown by Ref. \cite{2015AJ....150....7A,1997ApJ...478..745W}), so we adopted a double Gaussian fit. This approach will also be useful for future comparisons with the flare properties reported in Ref. \cite{2015AJ....150....7A}.}

Extended Data Figure 2 shows the FUV line profiles other than the C~\textsc{III} and Si~\textsc{IV} lines presented in Figure \ref{fig:fuv}. For the hydrogen Ly$\alpha$ line at 1215.7~\AA, we examined the intensity ratio between the blue and red wings across both broad and narrow wavelength intervals. In panel (b), we find a slight enhancement in the narrow blue wing, corresponding to velocities of approximately $-200$ to $-400$ km s$^{-1}$, which is consistent with the velocities observed in the C~\textsc{III} and Si~\textsc{IV} lines.
A recent study reports that blue-wing Ly$\alpha$ emission during solar flares is associated with filament eruptions \citep{2025arXiv250417667M}. Therefore, the observed slight asymmetry may support the presence of cool plasma eruptions in this event.
Similarly, the C~\textsc{II} and Si~\textsc{IV} 1402.8~\AA\ lines also show slight enhancements in their blue wings. However, caution is warranted in these cases due to potential blending with nearby lines.
N V does not show a significant blueshifted enhancement, but as in panel (d),
the blueshift signal level estimated from Si IV blueshifts and Si IV / N V flaring flux ratio, indicated with red dashed line, is below or comparable to the noise level. 
\Add{O V 1371.3 {\AA} line is also too weak to be used for the spectral profile analysis.}  
Even if a blueshifted component is present, it would be difficult to detect in these weaker lines.
Extended Data Figure 3 also presents the light curves and FUV spectra for the minor flares that occurred on March 30 and March 31, 2024.

\subsection*{H$\alpha$ Spectroscopic Data Analysis}

We organized a coordinated observation campaign involving three ground-based telescopes in Japan and South Korea, conducted simultaneously with HST. Fortunately, clear skies at all sites allowed us to successfully obtain time-resolved H$\alpha$ spectra with all three telescopes. 
We commonly processed the data by using \texttt{IRAF} and \texttt{PyRAF} packages. 

We conducted fiber-fed echelle spectroscopic observations using the Bohyunsan Optical Echelle Spectrograph \citep[BOES;][]{kim2007} on the 1.8m telescope at the Bohyunsan Optical Astronomy Observatory (BOAO), Korea. BOES covers a wavelength range of 3600–10500 \AA, and we used the 300 $\mu$m fiber, yielding a spectral resolution of \( R \sim 30,000 \). The observations were carried out over four nights, from March 29 to April 1, 2024, with an exposure time of 10 minutes and a readout time of 97 seconds. Data reduction was performed using \texttt{boespy}, which includes bias and dark subtraction, flat-fielding, Laplacian cosmic-ray removal, extraction of 1D spectra for each spectral order, and wavelength calibration incorporating both heliocentric and stellar line-of-sight velocity corrections. The wavelength calibration was performed using a ThAr lamp, with calibration files obtained before the observations.  

We also conducted spectroscopic observations of H$\alpha$ using the Medium And Low-dispersion Long-slit Spectrograph (MALLS) on the 2.0m Nayuta Telescope at the Nishi-Harima Astronomical Observatory, Japan. MALLS covers a wavelength range of 6350–6800 \AA\ with a spectral resolution of \( R \sim 10,000 \). Observations were conducted from March 29 to April 1, 2024, under favorable weather conditions. Data reduction followed the procedure described in \cite{2022NatAs...6..241N}, including overscan and dark subtraction, flat-fielding, and wavelength calibration. The Fe-Ne-Ar comparison lamp installed in MALLS was used for wavelength calibration, with calibration files obtained before the observations. Since MALLS exhibits slight wavelength shifts due to temperature variations during the night, additional wavelength calibration was performed using photospheric lines near H$\alpha$. 
The following specific procedure was applied to calibrate telluric line contamination: first, the profiles and variations of the telluric lines were modeled using nearby water vapor features; second, assuming similar temporal evolution, global trends in the water vapor lines, as indicated by red arrows in Extended Data Figure 4, were removed.
The Nayuta data are significantly affected by variations in water vapor lines. As a result, its dynamic spectrum is not included in the main text but is shown in Extended Data Figure 4 to provide supportive evidence for the blueshifted absorption, where contamination from telluric lines is considered minimal.

Additionally, we performed optical spectroscopic monitoring using the Kyoto Okayama Optical Low-dispersion Spectrograph with Optical-Fiber Integral Field Unit (KOOLS-IFU) \citep[][]{2019PASJ...71..102M} on the 3.8m Seimei Telescope at Okayama Observatory, Japan. 
KOOLS-IFU covers a wavelength range of 5800–8000 \AA\ with a spectral resolution of \( R \sim 2000 \). Observations were carried out from March 29 to April 1, 2024, with exposure times of 1 minute and a readout time of 17 seconds, under favorable weather conditions. Data reduction followed the methods described in \cite{2022NatAs...6..241N,2024ApJ...961...23N}, including the extraction of wavelength-calibrated and continuum-normalized 1D spectra for each frame.  

Each telescope provided data with different time and wavelength resolutions, each playing a distinct role in the interpretation of the results.  
In particular, the 1.8m BOAO telescope data had the highest wavelength resolution and was less affected by telluric lines. However, its time cadence was the lowest among the three telescopes. Given these characteristics, we primarily used the 1.8m BOAO telescope data for H$\alpha$ analysis.  
The 2.0m Nayuta telescope data confirmed the presence of blueshifted absorption (see Extended Data Figure 4) but was affected by variations in strong telluric line absorptions due to changes in the star’s elevation. While some telluric features were well-calibrated, others remained blended with H$\alpha$ and nearby stellar absorption lines, necessitating their removal. We consider that this contributed to the small differences observed in the H$\alpha$ light curves in Extended Data Figure 6. Consequently, the Nayuta data were used as supporting evidence for the detection of H$\alpha$ blueshifted absorption, but we did not fit the spectral shape, as variations in water vapor lines were expected to contribute to the spectra.  
The 3.8m Seimei telescope provided the highest time cadence among the three, but its spectral resolution was insufficient to capture the low-velocity and small blueshift observed in the BOAO and Nayuta data. Therefore, its use was limited to light curve analysis. Nevertheless, the Seimei data supported the BOAO light curve and confirmed that the FUV-bright flare exhibited weak H$\alpha$ emission. This suggests that the flare energy was below $\gtrsim 10^{33}$ erg, as all previously reported superflares with energies above this threshold exhibited significant H$\alpha$ emission with EW $\gtrsim 1.0$ {\AA}, just above the sensitivity limit.

Here we define positive equivalent width (EW) as emission. To calculate the equivalent width of the H$\alpha$ line, we used the continuum-normalized spectrum. The wavelength integration range was set to 6562.8$-10$ {\AA} to 6562.8$+10$ {\AA}. However, for the Nayuta data, noise was observed around 6562.8$-10$ {\AA}, so the integration range was adjusted to 6562.8$-7.5$ {\AA} to 6562.8$+7.5$ {\AA}.

\subsection*{\Add{Emission Mechanism of the Long-Duration H$\alpha$ Flare}}

\begin{figure}[!hb] 
\begin{center}
\includegraphics[width=1.0\linewidth,clip]{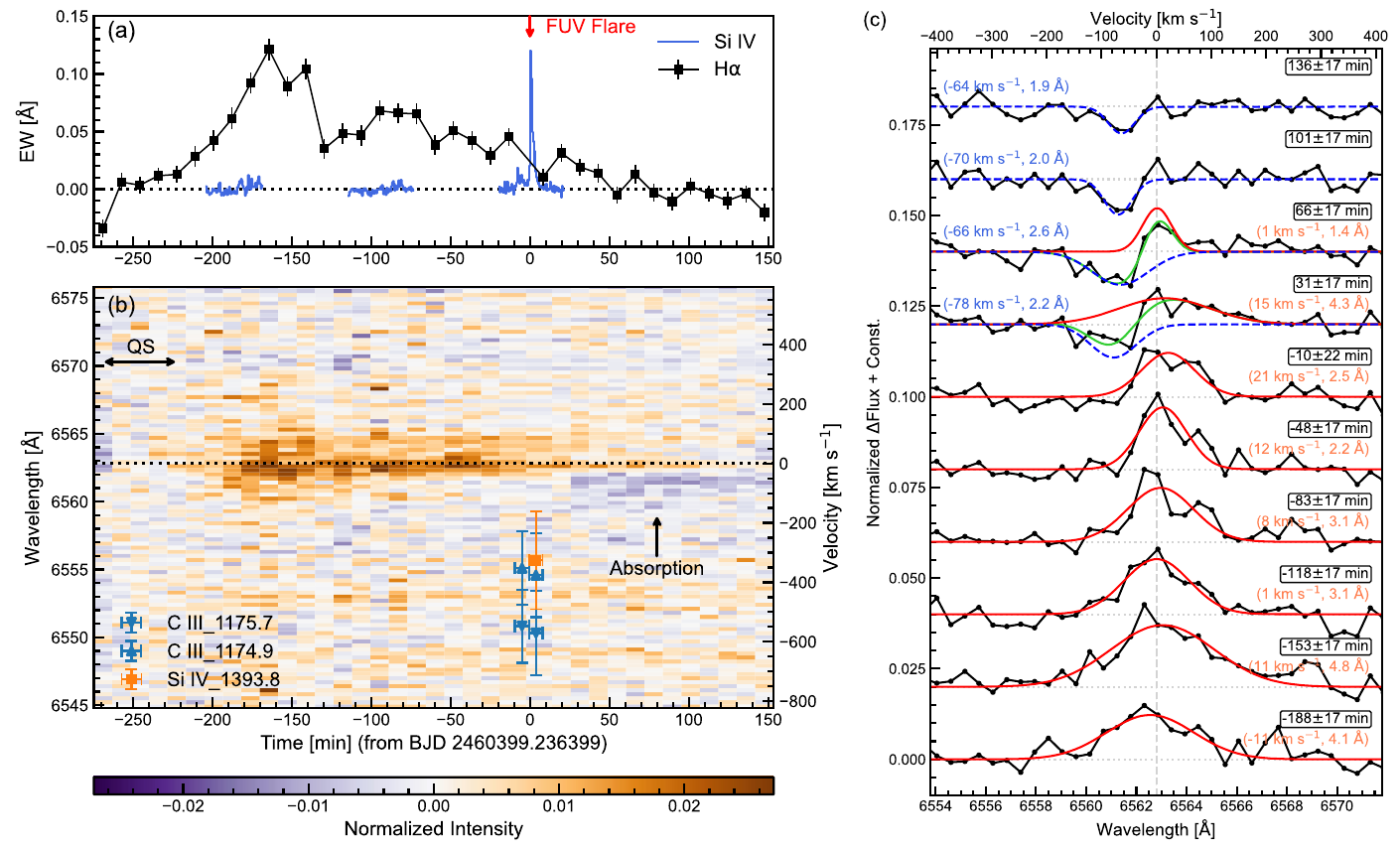}
\caption{H$\alpha$ spectrum exhibiting a blueshifted absorption component following the FUV flare.
(a) Light curve of the H$\alpha$ equivalent width (EW) in units of {\AA} (black squares), overplotted with the normalized light curve of Si IV 1393 {\AA} emission (blue line). 
The error bars of H$\alpha$ EW are calculated as the square root of the sum of squared standard deviations of the continuum spectrum relative to the template spectrum.
The FUV flare shown in Figure \ref{fig:fuv} is indicated in red. Notably, the H$\alpha$ line shows an earlier, long-duration flare beginning approximately 3 hours before the FUV flare and nearly decaying by the time the FUV flare occurs.
(b) Dynamic spectrum of the H$\alpha$ line after subtraction of the quiescent (pre-flare) spectrum, obtained during the period marked by black arrows and labeled ``QS." Orange and blue colors indicate excess emission and absorption, respectively, relative to the reference spectrum. The dotted line marks the central wavelength of the H$\alpha$ line. Blue and orange points denote the velocities of blueshifted components in the C III and Si IV lines, respectively. Note that the C III line is a blended multiplet; thus, velocities for both the strongest (1175.7 {\AA}) and bluest (1174.9 {\AA}) components are plotted using different symbols. The velocity error bars indicate the velocity dispersion derived from Gaussian fits to the blueshifted components while the time error bars are the integration time of spectra.
(c) Spectral fitting of the H$\alpha$ line. To improve the S/N, multiple frames were binned. The red and blue lines represent Gaussian fits to the emission and absorption components, respectively, while the green line shows their combined profile, fitted simultaneously. The central velocity and line width for each component are annotated.}
\label{fig:ha}
\end{center}
\end{figure}

\begin{figure}[!hb] 
\begin{center}
\includegraphics[width=0.75\linewidth,clip]{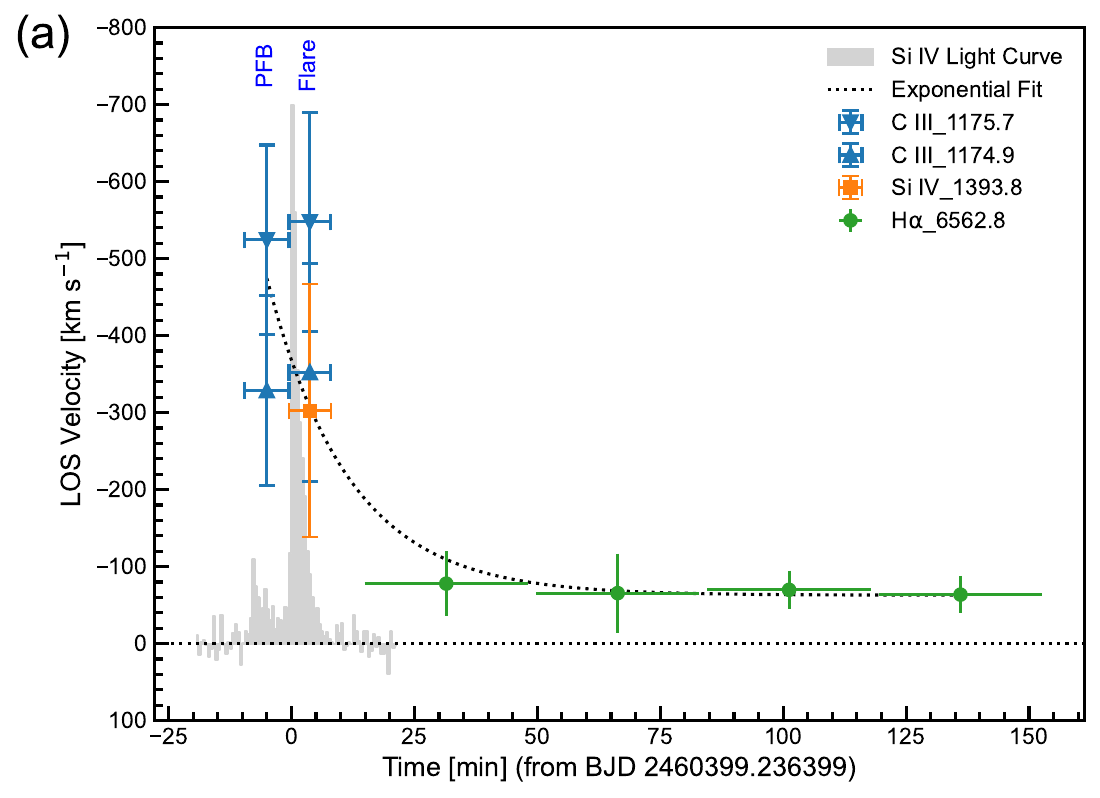}
\includegraphics[width=1.0\linewidth,clip]{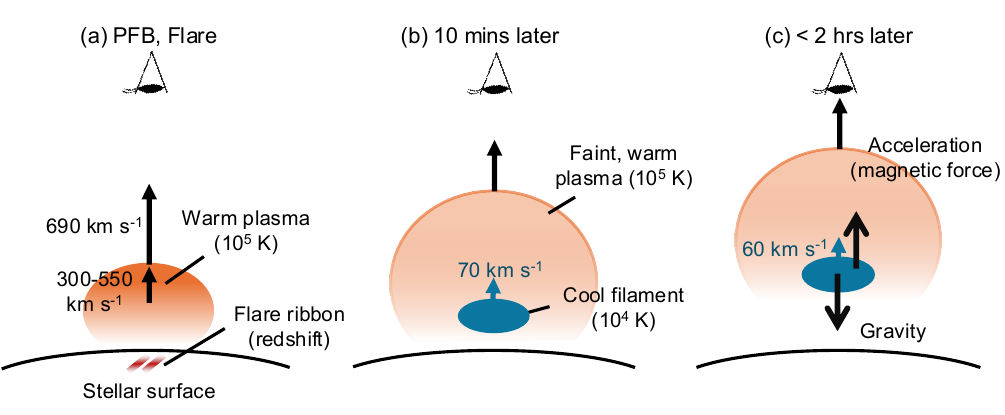}
\caption{Time evolution of blueshifted velocities in FUV and optical wavelengths.
(a) Each data point represents the central velocity derived from a Gaussian fit, where the vertical error bars correspond to the Gaussian standard deviation. 
The horizontal error bars indicate the time intervals over which the spectra were integrated for the fitting.
The C III line is a blended multiplet; therefore, velocities for both the strongest (1175.7 {\AA}) and bluest (1174.9 {\AA}) components are plotted using distinct symbols.
The time axis is referenced to the onset of the FUV flare, and the normalized Si IV light curve is shown as a shaded gray curve. 
The black dotted line indicates a simple exponential fit to the data.
(b) Schematic illustration of possible explanations for the time evolution of blueshifts in the FUV and H$\alpha$. The left panel depicts a warm plasma eruption occurring during the ``PFB" and ``Flare" phases. The middle and right panels illustrate scenarios relevant to the period $\sim$10 minutes after the flare, when only H$\alpha$ filament eruption starts to be observed. 
The middle panel represents the possibility of a multi-temperature, multi-layer structure within a single eruptive event. 
The right panel illustrates an alternative scenario involving multi-temperature eruptions at different locations, which are likely physically connected, analogous to ``sympathetic" events observed on the Sun.
Here, sympathetic eruptions means phenomena where a eruption caused by another eruption in a nearby or connected region.
In ten minutes later, the warm plasma became faint and undetectable in the FUV line wings, likely due to its expansion. However, it is illustrated in the panels as still present, either continuing to expand or influencing the surrounding environment.
}
\label{fig:vel}
\end{center}
\end{figure}

\Add{Here we briefly discuss the potential emission mechanism of the long-duration H$\alpha$-loud flare at -200$\sim$-150 min in Figure \ref{fig:ha}. 
The long-duration H$\alpha$ flare with almost no strong FUV enhancement is an interesting finding in our dataset, especially in contrast to the FUV-loud flare discussed in the main text. 
We briefly outline three possible interpretations for the lack of FUV emissions.
First is the geometric effect: the long-duration H$\alpha$-bright flare may have occurred just behind the limb, so that the flare ribbons (emitting in both H$\alpha$ and FUV) were invisible while only post-flare loops (emitting in H$\alpha$) were visible. 
However, the slight redshift seen in H$\alpha$ (Figure \ref{fig:ha}) makes this less likely, as both red- and blueshifted downflows from post-flare loops would be expected if occurring on the limb. 
Second is an exceptional Neupert-effect case: in solar flares, H$\alpha$ tends to track the thermal soft X-ray light curve, while FUV emission often follows the non-thermal hard X-ray profile, leading to a Neupert-effect-like relation \citep{2022ApJ...933..209N}. 
In this case, we would expect the FUV peak to occur during the rising phase of the H$\alpha$ flare, mostly within the first HST orbit. 
However, not all events clearly follow the Neupert effect, as also seen in M-dwarf flares \citep{2023ApJ...951...33T}. 
It is therefore possible that, due to an exceptional Neupert effect, the FUV peak occurred just before the first HST orbit and was missed in our observations. 
Third is a low heating-rate flare: if the heating rate per unit area is very low, non-thermal emissions such as FUV and white-light flares can be suppressed, while thermal emissions such as GOES X-rays and H$\alpha$ remain strong. Such events, known as ``non-white-light" flares, are often observed on the Sun \citep{2017ApJ...850..204W}.
Further details will be discussed in our upcoming paper.}

\subsection*{Flare Energy Calculation}

Bolometric white-light flare energies, spanning from UV to optical wavelengths, have served as a primary metric in stellar flare studies for over a decade. These energies are typically estimated from Kepler and TESS broadband photometry under the assumption of a 9,000–10,000 K blackbody spectrum \citep{2012Natur.485..478M}. To place our detected flares in this context, we estimated the bolometric UV–optical energy using the same blackbody approximation and compared the energy scale of our flares with those from previous statistical studies.
However, it is important to note that recent flare observations and models--primarily performed for M-dwarf and solar flares--indicate that the spectral energy distribution is not always well represented by a single blackbody (for review, please refer to Ref. \citep{2024LRSP...21....1K} and references therein).
Some flares are thought to include optically thin emission components, such as enhanced Balmer continuum in the UV \citep{2013ApJS..207...15K}, while others consist of multiple flare kernels \citep{2025ApJ...978...81K}. 
Furthermore, Ref. \cite{2022AJ....164..110F} \Add{and \cite{2019ApJ...871L..26F}} reported excess blue continuum emission at FUV wavelengths $\lesssim 1100$~{\AA} and a recent study by Ref. \cite{2025ApJ...978...81K} shows a similar excess in the blue continuum in the NUV, further highlighting the complexity of FUV flare spectra.
Thus, the nature of continuum emission across the FUV to optical range remains uncertain. As our aim here is to estimate the energy scale rather than to determine the emission mechanism, we adopt a simplified approach. Nonetheless, the possibility of lower optical depths in the continuum remains, \Add{which may affect the energy estimates \citep{2024MNRAS.528.2562S}}, and further investigation is required to obtain accurate energy estimates.

As in the main text, associated with the major FUV flare, TESS observations did not detect significantly large flares (Extended Data Figure 6), defined as events exceeding three times the photometric error in two consecutive points. 
We determined an upper limit of flaring enhancement as 0.022\% of the TESS-band quiescent luminosity, corresponding to $\lesssim 3.3\times 10^{25}$ erg s$^{-1}$ {\AA}$^{-1}$ at an effective wavelength of 7687.77 {\AA} for G-dwarfs.

Instead of using TESS-band photometric data, we estimated the bolometric white-light energy from the FUV continuum.
Extended Data Fig. 5 shows the continuum spectrum for the flare only and quiescence.
FUV continuum fluxes at their flare peak were used (all of the continuum band has the same peak timing with 30 s cadence data).
We fitted these spectra using a simple blackbody radiation model (Planck function) to derive their temperatures and radiation areas.
\begin{eqnarray}\label{eq:1}
    L_{\rm flare}(\lambda, T) &=& A_{\rm flare} \left( \pi B(\lambda, T_{\rm flare}) - \pi B(\lambda, T_{\rm star}) \right),
\end{eqnarray}
where $A_{\rm flare}$ is area, $T_{\rm flare}$ is flare temperature, $B(\lambda, T)$ is Planck function, and $T_{\rm star}$ is stellar effective temperature.
\Add{Here, in this description, if the chromosphere during a flare can depart from LTE condition due to the high excitation, one may assume that the Equation \ref{eq:1} requires a correction for departure coefficient $b$ (i.e., $B(\lambda, T_{\rm flare})/b$). However, radiative hydrodynamic flare simulations with RADYN generally show that, in most cases, the high collisional rates due to compression and ionization drive the chromospheric condensation regions toward almost LTE, so that cases with $b\ll 1$ are rarely found in practice \cite{2015SoPh..290.3487K,2017ApJ...836...12K}. Therefore, while our blackbody assumption could introduce systematic uncertainties at the level of a factor of a few, we do not expect order-of-magnitude errors in the bolometric energy estimate from this point.}

The function \texttt{curve\_fit} was used for the fitting.  
As a result, the flare temperature $T_{\rm flare}$ is estimated to be 12,900$\pm$500 K and the area $A_{\rm flare}$ is $1.8\pm 0.6\times 10^{18}$ cm$^2$ (= $67 \pm 22$ millions of stellar hemisphere).
Extending the fit of the FUV continuum (with a temperature of 12,900 K) to the TESS band is almost consistent with the observed upper limit.
Based on the flaring temperature, the bolometric white-light luminosity is estimated to be $2.8 \pm 1.1 \times 10^{30}$ erg s$^{-1}$, and the estimated energy is $2.3 \pm 1.1 \times 10^{32}$ erg (X23 in GOES classification).
\Add{The estimated area is much smaller than the typical areas of ``superflares" ($\gtrsim$10$^{33}$ erg) on solar-type stars observed by Kepler \citep{2012Natur.485..478M}, which have a median of $\sim$1200 MSH.}
As a reference, the well-known solar Carrington event (often estimated to be around an $\sim$X45-class flare \citep{2022LRSP...19....2C}, although there are various interpretations \citep{2023ApJ...954L...3H}) is estimated to have had a white-light flare emitting area of 116$\pm$25 millionths of the solar hemisphere (MSH) on the basis of his sketch \citep{2023ApJ...954L...3H}.   
As alternative reference, we analyzed the IRIS's slit-jaw image at the flare peak of the X9.0-class solar flare in Extended Data Figure 7 (see Section ``Solar Data Analysis") and estimated the flaring area of its FUV emission line (primary C II doublet) as \Add{1.5$\times 10^{18}$ cm$^{2}$ = 50 MSH}.
\Add{We also examined the SDO/HMI continuum and AIA 1600 {\AA} data and confirmed that, at the flare peak, most of the flare ribbon lies within the IRIS FoV, and that the SDO/HMI continuum flare ribbon area of $\sim$25 MSH is broadly consistent with the above estimates.}
Our estimate of the flaring area is roughly consistent with that of solar $\sim$X10-class flares, supporting the reasonableness of our estimates for both energy and area. 
\Add{As noted in the main text, such X10-class flares can be detected with TESS on low-luminosity M, K, or late G-dwarfs \citep[e.g.,][]{2023ApJ...954...19P}, but are mostly difficult to detect on brighter early G-dwarfs like EK Dra due to their relatively small flux increase.
}

As an estimate from another solar-stellar comparison, we calculated the Si IV 1393.8 {\AA} flaring luminosity to be $(1.8 \pm 0.1) \times 10^{28}$ erg s$^{-1}$ and the total energy to be $(1.8 \pm 0.1) \times 10^{30}$ erg for the FUV flare observed on EK Dra.
This is compared with the Sun-as-a-star empirical model known as the Flare Irradiance Spectral Model Version 2 (FISM2), which provides solar spectral irradiance from 0.01 to 190 nm at 0.1 nm resolution with a 60 s cadence. The FUV component of FISM2 is based on SORCE/SOLSTICE observations, while the GOES/XRS-B flux and its derivative are used as proxies to reconstruct flare enhancements.
We selected a well-known X17-class solar flare that occurred on 28 November 2003--one of the so-called “Halloween” events known for their strong geoeffectiveness. The flux was integrated within ±1 {\AA} of the line center to construct the flare light curve. The Si IV 1393.8 {\AA} luminosity and energy for this event were estimated to be $(1.2 \pm 0.2) \times 10^{28}$ erg s$^{-1}$ and $(1.5 \pm 0.1) \times 10^{30}$ erg, respectively.
These values are very similar to those obtained for the FUV flare on EK Dra, supporting our estimate that the event corresponds to a flare of approximately X23-class.

\Add{In addition, we have performed a comparison with solar flares using the SDO/EVE flare catalogue (\url{https://lasp.colorado.edu/eve/data_access/eve-flare-catalog/index.html}) (a similar comparison between solar/stellar flares is presented in Ref. \cite{2018ApJ...867...71L}). 
By using solar flares whose locations are $<$45 deg from the disc center, we derived a correlation between the GOES X-ray peak flux and the C III 977 {\AA} radiation energy, which serves as a good proxy for C III 1175 {\AA} \cite{2018ApJ...867...71L}. 
This correlation is shown in Supplementary Fig. 1. 
From the C III 1175 {\AA} energy of the EK Dra flare, 2.1$\times 10^{30}$ erg, we estimated the corresponding C III 977 {\AA} energy by multiplying the emissivity ratio C III 977/C III 1175 = $\sim$1.5 at the electron density of 10$^{11-12}$ cm$^{-3}$ (obtained from CHIANTI \citep{2024ApJ...974...71D}).
And then, using the solar scaling, we derived a GOES class of X74$_{-54}^{+151}$ for the EK Dra flare.
This value is of the same order as the Carrington flare ($\sim$X45) and our estimates from the FUV continuum ($\sim$X23).}

\subsection*{Solar Data Analysis}

IRIS observes the Sun at high spatial, temporal, and spectral resolution in the NUV and FUV. We retrieved the Level 2 calibrated data of the X9 flare from the Lockheed Martin Solar Astrophysics Laboratory archive. The data were obtained as part of a high-cadence flare program (OBSID 4204700237) in sit-and-stare mode with a 0.8~s spectral cadence. 
Slit-jaw images at 1330 \AA\ (SJI 1330) were obtained with an 8~s cadence. The spatial and spectral directions employ binning by 2 pixels, thus giving a linear spatial dispersion of 0.334 arcsec pix$^{-1}$ and a linear spectral dispersion of 0.02596 \AA\ pix$^{-1}$ (FUV) and 0.0509 \AA\ pix$^{-1}$ (NUV). Spectral coverage spans 4 wavelength windows: $\lambda = 1332.7 - 1336.8$ \AA\ (centered on C II 1336), 1400.3 -- 1406.3 \AA\ (centered on Si IV 1403), 2793.6 -- 2800.0 \AA\ (centered on Mg II k 2796), and 2813.3 -- 2816.3 \AA\ (centered on Fe II 2814). Level 3 data are produced using the standard IDL routines. Inspection of the fiducial marks suggests, if any, a sub-pixel relative spatial offset between the FUV and NUV channels; thus we do not shift the spectra during generation of the Level 3 files. Radiometric calibration of the spectra and slit-jaw images from corrected DN s$^{-1}$ to erg cm$^{-2}$ s$^{-1}$ sr$^{-1}$ \AA$^{-1}$ follows IRIS Technical Note 24 (\url{https://iris.lmsal.com/documents.html}), accounting for the binning factors and the time-dependent sensitivity with \texttt{iris\_get\_response.pro} (v005). According to ITN 22 and Section 1.3 of ITN 32, the header keywords for XCEN and YCEN may be inaccurate to about $\pm 5$ arcsec. Thus, we manually coalign the IRIS spectra and slit-jaw images with SDO data. Using a shift of $-3.5$ arcsec in IRIS $x$ and a $-3.$ arcsec shift in IRIS $y$, we find that the flare ribbon features in SJI 1330 at UTC 12:13:49.5 align with the same features in SDO/AIA 1600 at UTC 12:13:50.1 (before the SDO data saturate).

Extended Data Figure 7(a) shows the slit-jaw (SJ) image at 1330~\AA, taken on 3 October 2024 at 12:17:31.547 UTC. The dark vertical line indicates the slit position, which intersects the two flare ribbons. Although the IRIS slit does not cover the entire flare region, we can simulate Sun-as-a-star-like flare properties by integrating the intensity over the two ribbons intersected by the slit.
Extended Data Figures 7(b) and (c) compare the Si~\textsc{iv} and C~\textsc{ii} spectra from three datasets: the spatially-averaged solar flare on the slit, the spatially and temporally-averaged solar flare spectrum, and the EK Dra flare. 
All spatially integrated spectra show gentle redshifts with velocities of $\leq 50$ km s$^{-1}$, which are commonly interpreted as signatures of chromospheric condensation or coronal rain. Notably, while the solar flare spectra exhibit no extended blue-wing emission, the EK Dra flare shows a prominent extended blue wing.
\Add{Note that while the difference between solar/stellar data in the Si IV red wing is small, the C II doublet (1334.53 {\AA} and 1335.70 {\AA}) shows a more noticeable difference. 
The red wing of C II 1334.53 {\AA} in the HST stellar spectra is strongly absorbed by the interstellar medium at around 1334.62 {\AA} \citep{2015AJ....150....7A}. This results in the apparently stronger redshifts for C II 1334.53 {\AA} in the solar profile.
For the long-wavelength line at 1335.70 {\AA}, the strong redshifts in the solar data are real and noteworthy, but this line is not the primary focus of the present study and will be analyzed in our forthcoming paper.
}

\subsection*{Plasma Mass Estimation}

The luminosity of the blueshifted component in the C~\textsc{iii} and Si~\textsc{iv} lines is $9.8 \pm 3.6 \times 10^{26}$ erg s$^{-1}$ and $1.7 \pm 0.5 \times 10^{27}$ erg s$^{-1}$, respectively. From this, we estimated the mass of the warm eruptive filament, assuming that both lines are optically thin. While C~\textsc{iii} is not necessarily optically thin, we adopt this assumption considering the expanding filament and its filling factors.
Using CHIANTI ver.11 \citep{2024ApJ...974...71D}, we computed the emissivity for each line under optically thin conditions, assuming solar coronal abundances.

We calculated the emissivity for electron densities in the range of $\log n_e$ [cm$^{-3}$] = 9.0–12.0 in steps of 0.1. The electron density at which the modeled line ratio of C\ \textsc{iii} to Si\ \textsc{iv} matches the observed ratio was found to be $\log n_e = 9.3$. Based on this value, we derived an emission measure of $EM_{\rm Si\_IV} = (4.8 - 9.0) \times 10^{49}$ cm$^{-3}$, a plasma mass of $M_{\rm Si\_IV} = (4.0 - 7.6) \times 10^{16}$ g, and a kinetic energy of $E_{\rm kin, Si\_IV} = (1.1 - 4.9) \times 10^{31}$ erg, assuming a blueshifted velocity of $V_{\rm Si\_IV} = -300 \pm 60$ km s$^{-1}$.
Here, subscripts such as ``$\rm C\_III$" and ``$\rm Si\_IV$" indicate that the corresponding physical quantities were derived specifically for the C~\textsc{III} and Si~\textsc{IV} lines, respectively.
These estimates are based primarily on the Si\ \textsc{iv} line, which has less ambiguity in the velocity measurement.
The inferred electron density is reasonable in the context of solar filament eruptions of 10$^{9-11}$ cm$^{-3}$ \citep{2014LRSP...11....1P}. Previous X-ray observations of EK Dra report coronal densities of $\log n_e$ [cm$^{-3}$] = 10.5–11.5 \citep{2004A&A...427..667N}. Considering that the plasma is likely expanding, the lower estimated density remains plausible.
\Add{Here we note that the emission measure from evaporation-origin blueshifts can be up to $\sim1.8 \times 10^{47}$ cm$^{-3}$ when we assume transition-region density of 10$^{10}$–10$^{11}$ cm$^{–3}$, a thickness of 10$^7$–10$^8$ cm, and an upper-limit case in which 10\% of the flaring area (1.8$\times$10$^{17}$ cm$^2$) shows blueshifts, which is unlikely to explain the observed emission measure.} 

We also estimated the plasma parameters assuming higher electron densities in the range of $\log n_e$ [cm$^{-3}$] = 10.5–11.5. This is because the C\ \textsc{iii} and Si\ \textsc{iv} emission may not originate from the same plasma, and uncertainties in their luminosity ratio prevent precise constraints on the electron density.
Based on this, the emission measure ($EM$) was determined to be $EM_{\rm C\_III} = (2.9 - 6.4) \times 10^{49}$ cm$^{-3}$ and $EM_{\rm Si\_IV} = (3.8 - 8.2) \times 10^{49}$ cm$^{-3}$, and the mass to be $M_{\rm C\_III} = (0.15 - 3.4) \times 10^{15}$ g and $M_{\rm Si\_IV} = (0.20 - 3.8) \times 10^{15}$ g.
These values are well consistent with each other, supporting the possibility that these blueshifts originate from the same plasma.
Based on the observed velocity ($V_{\rm C\_III}=-350 \pm 50\sim-550 \pm 50$ km s$^{-1}$ and $V_{\rm Si\_IV}=-300 \pm 60$ km s$^{-1}$), we can calculate the kinetic energy of $E_{\rm kin,~C\_III} =(0.69-61)\times 10^{29}$ erg and $E_{\rm kin,~Si\_IV} =(0.59-28)\times 10^{29}$ erg.

For the cool filament eruption, we followed the method of \cite{2022NatAs...6..241N} to estimate the length scale, mass, and kinetic energy of filament eruptions based on the absorption EW and velocity. The absorption EW is larger before \( t = 66 \pm 17 \) min, but this measurement is likely affected by blending with the emission component. To ensure a conservative estimate, we adopted an absorption EW of $0.013 \pm 0.005$ {\AA} and a velocity of $70 \pm 7$ km s\(^{-1}\) from the spectrum at \( t = 101 \pm 17 \) min. The impact of this choice is estimated to be within a factor of 2.
For reference, the absorption EW of a stellar eruptive filament on EK Dra, as reported in \cite{2022NatAs...6..241N}, was 0.16 {\AA}, an order of magnitude larger than that in this work. This clearly indicates that the filament eruption observed here is much smaller in scale.

We briefly outline the method used to estimate filament properties. We applied Becker's cloud model, assuming an optical depth of 5 (with an error range of 0.8–10), a two-dimensional aspect ratio of 1 (or 0.1), a dispersion velocity of 20 km s\(^{-1}\) (or 10–220), and a source function of 0.1 (or 0.02–0.5), based on solar observations. The modeled EW for the cloud was derived, and the observed stellar EW was divided by this value to estimate the filament area (i.e., the filament filling factor is given by observed EW/modeled EW). 
The velocity was used to calculate the typical modeled EW, considering the wavelength dependence of absorption in the background H$\alpha$ profile of a solar-type star. From the estimated line-of-sight (LoS) length scale, we determined the hydrogen column density and calculated the total volume by multiplying the area by the column density and the hydrogen atom mass. Finally, the kinetic energy was estimated using the observed velocity and calculated mass. For further details, refer to \cite{2022NatAs...6..241N}.
As a result, we obtained a filament mass of $(0.93-31) \times 10^{16}$ g, a length scale of $(0.36-14) \times 10^{10}$ cm, and a kinetic energy of $(0.19-9.0) \times 10^{30}$ erg. The uncertainties reflect the errors in EW and velocity, as well as the assumed model parameter range.

To put the detected eruptions in the context of solar and stellar observations, the mass and kinetic energy as a function of flare radiation energy are plotted in Extended Data Figure 8. 
\Add{In the panel, to enable direct comparison between solar and stellar data, we adopt the scaling relations $E_{\rm bol} = 100 \times E_{\rm GOES}$ \citep{2012ApJ...759...71E} and $E_{\rm bol}$ [erg] = $10^{35} \times F_{\rm GOES}$ [W m$^{-2}$] \citep{2013PASJ...65...49S,2016A&A...588A.116W,2022LRSP...19....2C}.
Both are empirical solar equations, and although there is uncertainty in the energy distribution across different wavelengths, they have been shown to perform reasonably well in solar flare studies \citep{2012ApJ...759...71E,2016A&A...588A.116W, 2022LRSP...19....2C} and in some stellar flare studies \citep{2024ApJ...961...23N}.}

\clearpage



\section*{Data Availability}
The source data underlying the figures of this study are provided with this paper.
The raw spectroscopic data for ground-based observations are available either in the associated observatory archive \url{https://smoka.nao.ac.jp/index.jsp} or upon
request from the corresponding author. The TESS and HST data are available at the MAST archive (\url{https://mast.stsci.edu/portal/Mashup/Clients/Mast/Portal.html}). The IRIS data are available at the Virtual Solar Observatory (\url{https://sdac.virtualsolar.org/cgi/search}) or the Lockheed Martin Solar Astrophysics Laboratory archive (\url{https://iris.lmsal.com/search/}).

\section*{Code Availability}
Code availability: The basic codes for the HST data analysis are available on the following website \url{https://www.stsci.edu/hst/instrumentation/cos/documentation/notebooks}.
We followed the spectral analysis methods described on the following website \url{https://spacetelescope.github.io/hst\_notebooks/notebooks/COS/LSF/LSF.html} (version 2023-06-29).

\section*{Acknowledgements}

We are grateful to the referees for their helpful suggestions that improved the quality of the paper.
This work was supported by the following fundings: JSPS KAKENHI Grant Numbers, JP24H00248 (K.N., H.M.), JP24K00680 (K.N., H.M., K.S.), JP24K00685 (H.M.), and JP25K01041 (K.N., H.M.); the Operation Management Laboratory of NINS OML022403 (K.N.); GSFC Sellers Exoplanet Environments Collaboration (SEEC), which is funded by the NASA Planetary Science Division’s Internal Scientist Funding Model (ISFM), NASA's Astrophysics Theory Program grant \#80NSSC24K0776NNH21ZDA001N-XRP, F.3 Exoplanets Research Program grants, NICER Cycle 2 project funds (V.S.A.); NASA TESS Cycle 6 Program 80NSSC24K0493 and NASA ADAP 80NSSC21K0632 (Y.N.); GSFC Internal Scientist Funding Model program (P.R.Y.). The research of the SNU team was supported by the National Research Foundation of Korea (RS-2023-00208117).
Spectroscopic data were obtained through the BOAO telescope, the Nayuta telescope (program 2311-T-10 of The Optical and Infrared Synergetic Telescopes for Education and Research) and the Seimei telescope (24A-N-CT08). 
This paper includes data from TESS and HST, archived at MAST (STScI), funded by NASA.
The HST-COS data were obtained as part of HST GO program 17595; support was provided by the associated grant to the University of Colorado at Boulder (K.F.).
IRAF is distributed by the National Optical Astronomy Observatories, which are operated by the Association of Universities for Research in Astronomy, Inc., under cooperate agreement with the National Science Foundation. 
PyRAF is part of the stscipython package of astronomical data analysis tools and is a product of the Science Software Branch at the Space Telescope Science Institute.

\section*{Author Contributions}

K.N., K.F., and J.C. led the campaign observations at their respective observatories.
K.N. performed the data analysis and mainly wrote the manuscript.
K.N., K.F., J.C., S.H., S.K., J.K., K.L., T.O., and M.T. contributed to the observations.
A.K., C.T., and K-S.L. contributed to the analysis of solar data.
P.Y. contributed to the CHIANTI modeling.
K.N., K.F., J.C., V.A., A.K., Y.N., H.M., and K.S. contributed to the theoretical interpretation.
All authors reviewed and commented on the manuscript.

\section*{Competing Interests}

The authors do not have any conflicts of interest to report. 




\section*{Figures}

\renewcommand{\figurename}{\bf\noindent{Extended Data Fig.}}
\setcounter{figure}{0}


\begin{figure}
\begin{center}
\includegraphics[width=1.0\linewidth,clip]{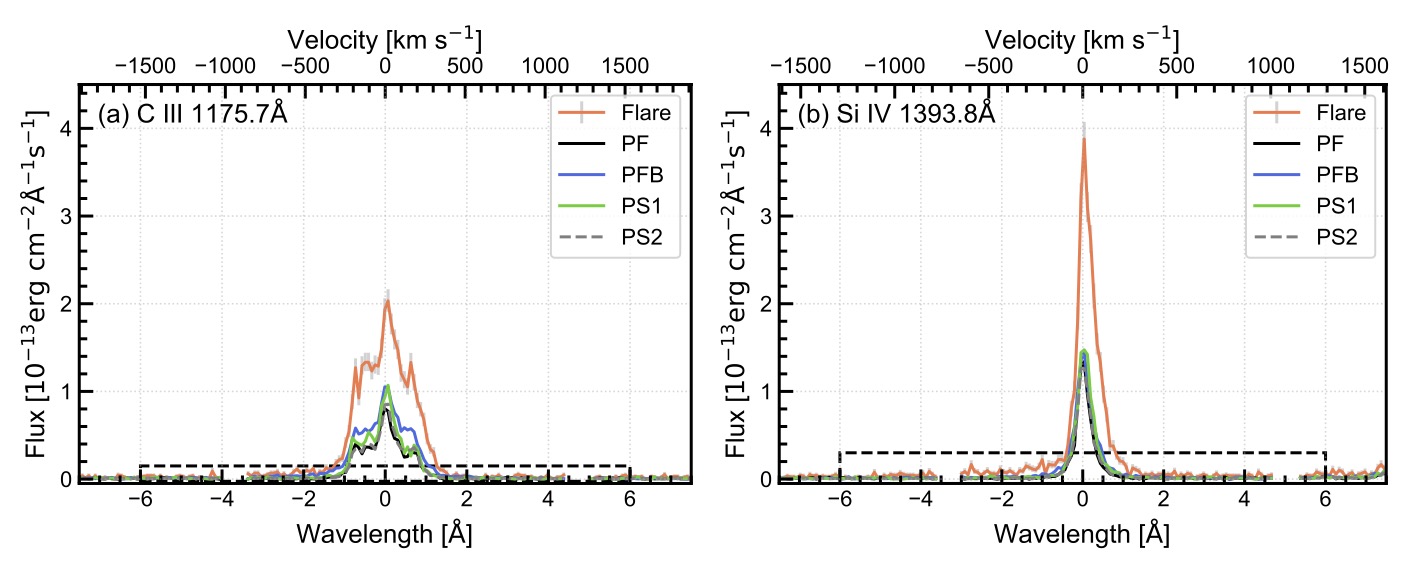}
\caption{Whole flaring profiles for C III and Si IV lines associated with flares. The spectra were obtained for five phases--pre-flare (PF), pre-flare brightening (PFB), flare, post-flare 1 (PS1), and post-flare 2 (PS2).
The plotted data are the same spectra as Figure \ref{fig:fuv}, but are shown with a much larger flux and wavelength range to illustrate the overall spectral shape and line blending. 
A black dashed outline indicated the x-y range of Figure \ref{fig:fuv}. As a result, the blue-wing emission appears less visible in this figure, but Figure \ref{fig:fuv} shows that it is indeed significant above the flux errors.
}
\label{fig:ext}
\end{center}
\end{figure}

\begin{figure}
\begin{center}
\includegraphics[width=1.0\linewidth,clip]{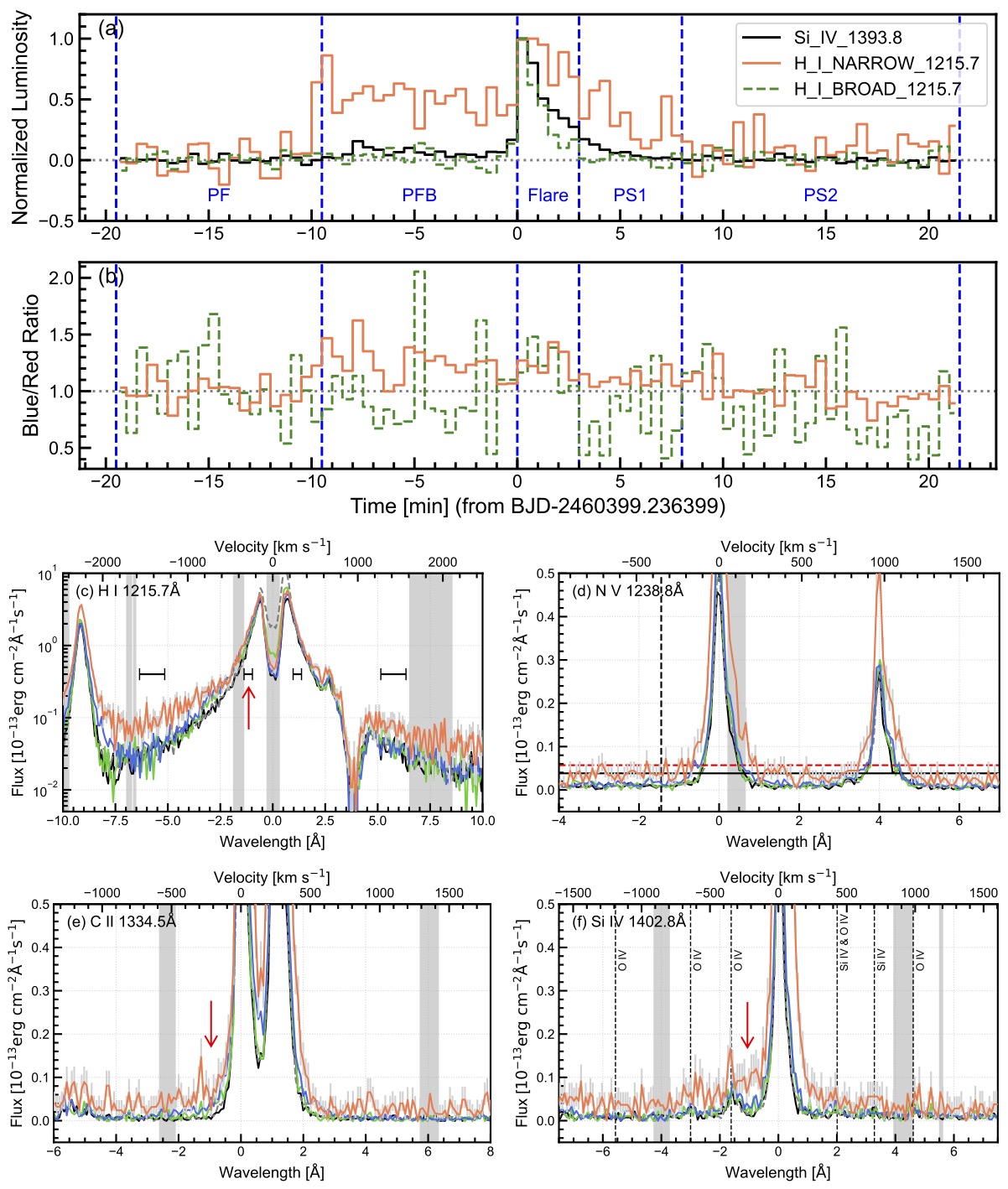}
\caption{FUV spectral lines associated with flares other than the one shown in Figure~\ref{fig:fuv}.
(a) Light curves of the narrow and broad wing components of the hydrogen Ly$\alpha$ line, as the wavelength ranges are defined as the black error ranges in panel (c).
(b) Time evolution of the blue-to-red wing flux ratio for both the narrow and broad wings for Ly$\alpha$ line.
(c–f) Temporal evolution of the spectral line profiles for hydrogen Ly$\alpha$, N V 1238 {\AA}, the C II doublet, and Si IV 1402 {\AA}, respectively. 
The gray shaded regions indicate wavelength ranges flagged with bad Data Quality (DQ) in the HST dataset.
N V does not show a significant blueshifted enhancement. 
In panel (d), the black vertical line indicates the velocity of the Si IV blueshift from Figure \ref{fig:fuv}. The black horizontal line represents the continuum level, measured by integrating over the range 1238.8 + (1–3) {\AA}. The red dashed line shows the estimated level of the continuum plus a possible blueshift component, calculated by scaling the Si IV blueshift signal by the N V / Si IV flaring flux ratio. Since the red dashed line is below or comparable to the noise level, even if a blueshifted component is present, it would be difficult to detect in these weaker lines.
C II appears to exhibit an extended blue wing component; however, this region is likely contaminated by a nearby blended emission line and affected by poor data quality on the blue side, requiring cautious interpretation.
Si IV 1402 {\AA} also shows an extended blue wing, but due to blending with O IV lines, this feature was excluded from the velocity analysis.
\Add{The red arrows indicate the possible locations of the blueshifted emission components.}
}
\label{fig:otherlines}
\end{center}
\end{figure}

\begin{figure}
\begin{center}
\includegraphics[width=1.0\linewidth,clip]{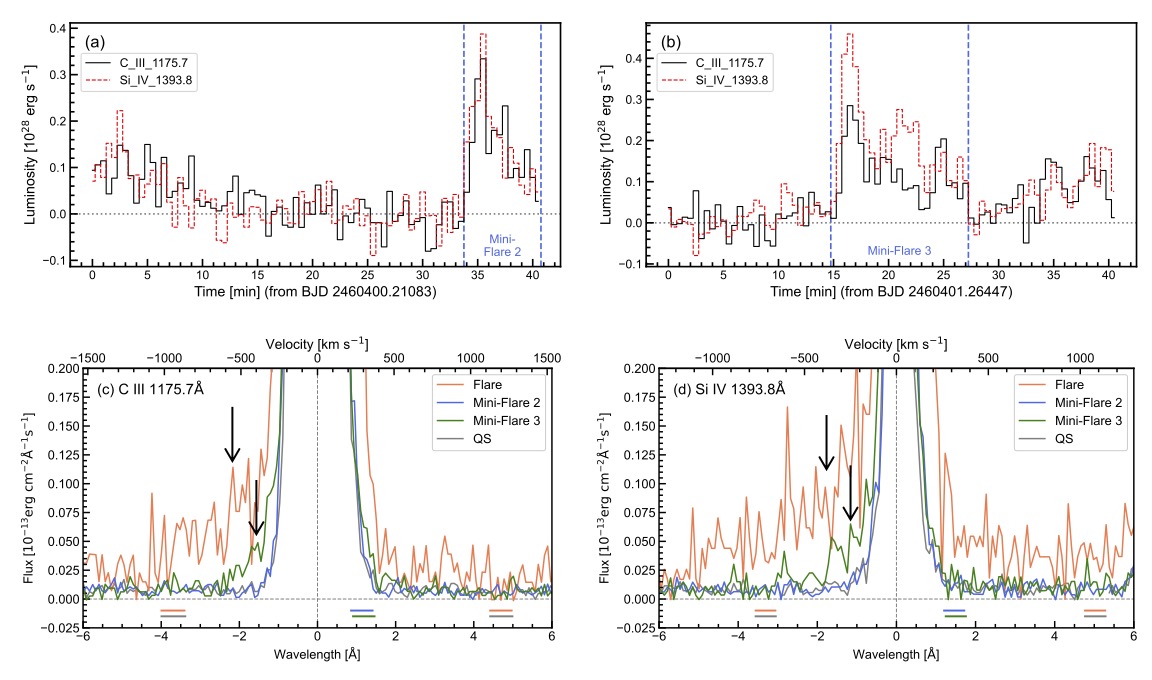}
\caption{FUV light curves and spectra of two minor flares (labeled Mini-Flares 2 and 3) observed with HST. (a, b) Light curves of C III (black solid lines) and Si IV lines (red dashed lines) for Mini-Flares 2 and 3. The time interval enclosed by the blue dashed lines is defined as the flare period, from which the spectra in panels (c) and (d) were extracted. (c, d) Spectral line profiles for C III and Si IV lines. The horizontal line with each color indicate the wavelength ranges affected by bad Data Quality (DQ) flags for each corresponding spectrum.
The black arrows indicate the approximate locations of the blueshifted emission components. 
}
\label{fig:miniflare}
\end{center}
\end{figure}

\begin{figure}
\begin{center}
\includegraphics[width=1.0\linewidth,clip]{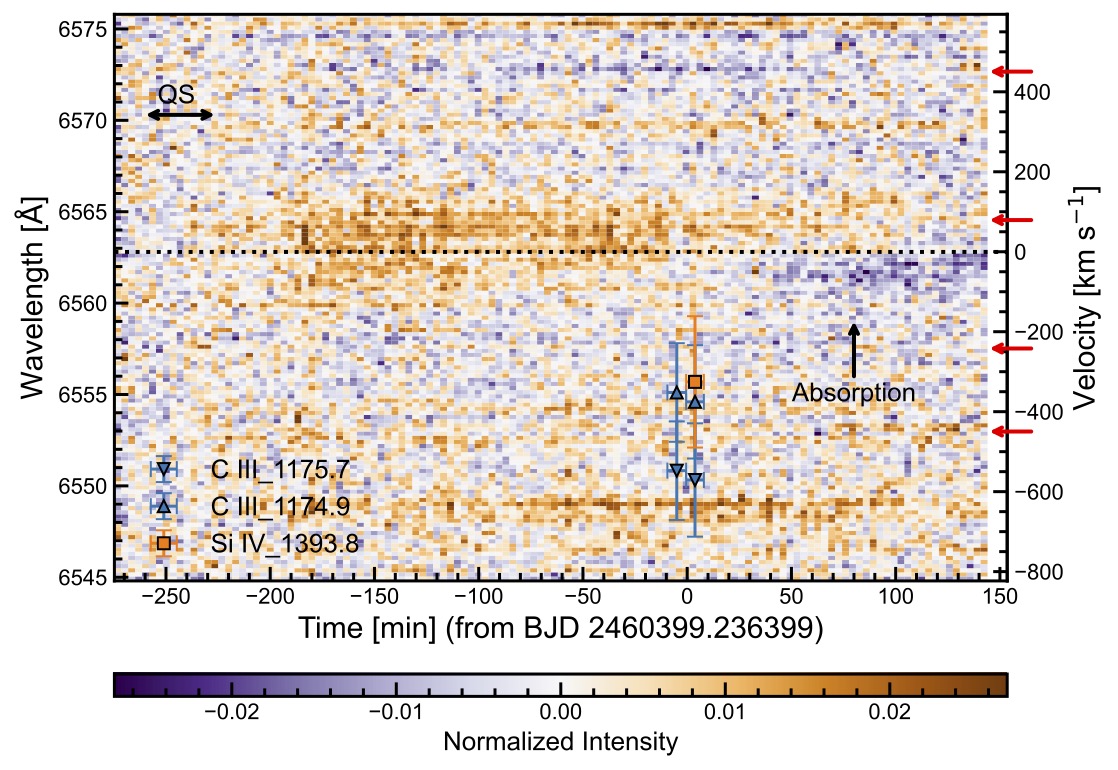}
\caption{Dynamic spectrum of the H$\alpha$ line observed with the Nayuta Telescope \Add{after subtraction of the quiescent (pre-flare) spectrum, obtained during the period marked by black arrows and labeled ``QS." Orange and blue colors indicate excess emission and absorption, respectively, relative to the reference spectrum. The dotted line marks the central wavelength of the H$\alpha$ line. Blue and orange points denote the velocities of blueshifted components in the C III and Si IV lines, respectively. Note that the C III line is a blended multiplet; thus, velocities for both the strongest (1175.7 {\AA}) and bluest (1174.9 {\AA}) components are plotted using different symbols.}
Red arrows indicate wavelengths affected by strong telluric water vapor lines and the trends at these wavelengths were removed using the method described in the text. 
}
\label{fig:nayuta}
\end{center}
\end{figure}

\begin{figure}
\begin{center}
\includegraphics[width=1.0\linewidth,clip]{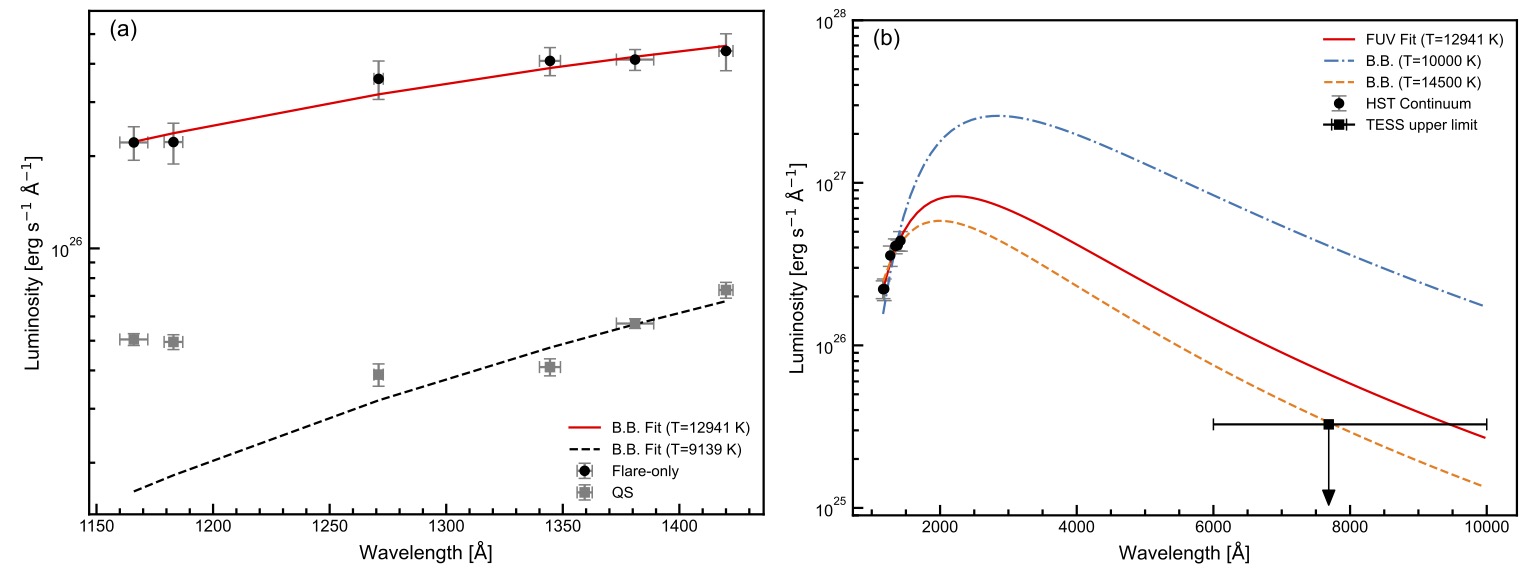}
\caption{Potential continuum spectrum of the FUV flare. (a) Blackbody fits for the flaring and the quiescent continuum. The extra emission in quiescent phase  below 1200 {\AA} could be a continuum from S I $\lambda < 1199$ {\AA}.
(b) The extension of the blackbody spectrum to the TESS band where we could not detect significant white-light emission.
\Add{All Planck curves are normalized to match the observed FUV continuum luminosity at 1160–1420 {\AA}. As a result, the higher-temperature curves (orange) are divided by a larger factor, leading to lower values at optical wavelengths. The errors in each flux are Poisson errors, as described in the text, while the wavelength errors are the corresponding wavelength ranges.}
}
\label{fig:cont}
\end{center}
\end{figure}

\begin{figure}
\begin{center}
\includegraphics[width=1.0\linewidth,clip]{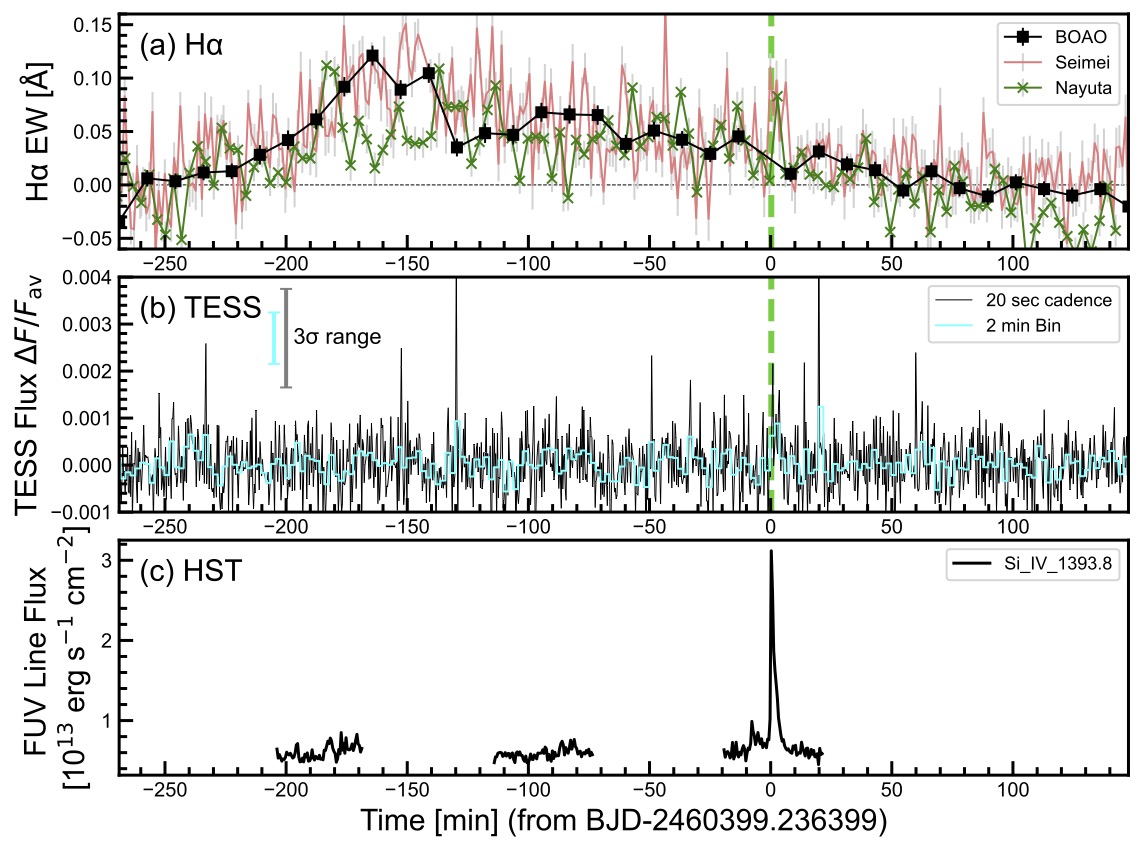}
\caption{FUV and optical light curves of EK Dra on 29 March 2024.
(a) H$\alpha$ light curves obtained with the 3.8m Seimei Telescope (red), the 2.0m Nayuta Telescope (green), and the 1.8m BOAO Telescope (black). The green dashed line indicates the timing of the UV flare. 
The horizontal axis shows time in minutes, measured from the onset of the largest FUV flare.
The error bars of H$\alpha$ EW are calculated as the square root of the sum of squared standard deviations of the continuum wavelengths of each spectrum relative to the template spectrum.
(b) TESS light curve with 20-second cadence (black) and the same data binned to 2-minute intervals (cyan). The 3$\sigma$ error range of detrended flux is represented by the symbols in the upper left corner (note that the difference between the upper and lower error bars corresponds to 3$\sigma$, and therefore the median has no particular meaning in this context).
(c) Light curve of the Si IV 1393.8 {\AA} line.
}
\label{fig:lcall}
\end{center}
\end{figure}

\begin{figure}
\begin{center}
\includegraphics[width=1.0\linewidth,clip]{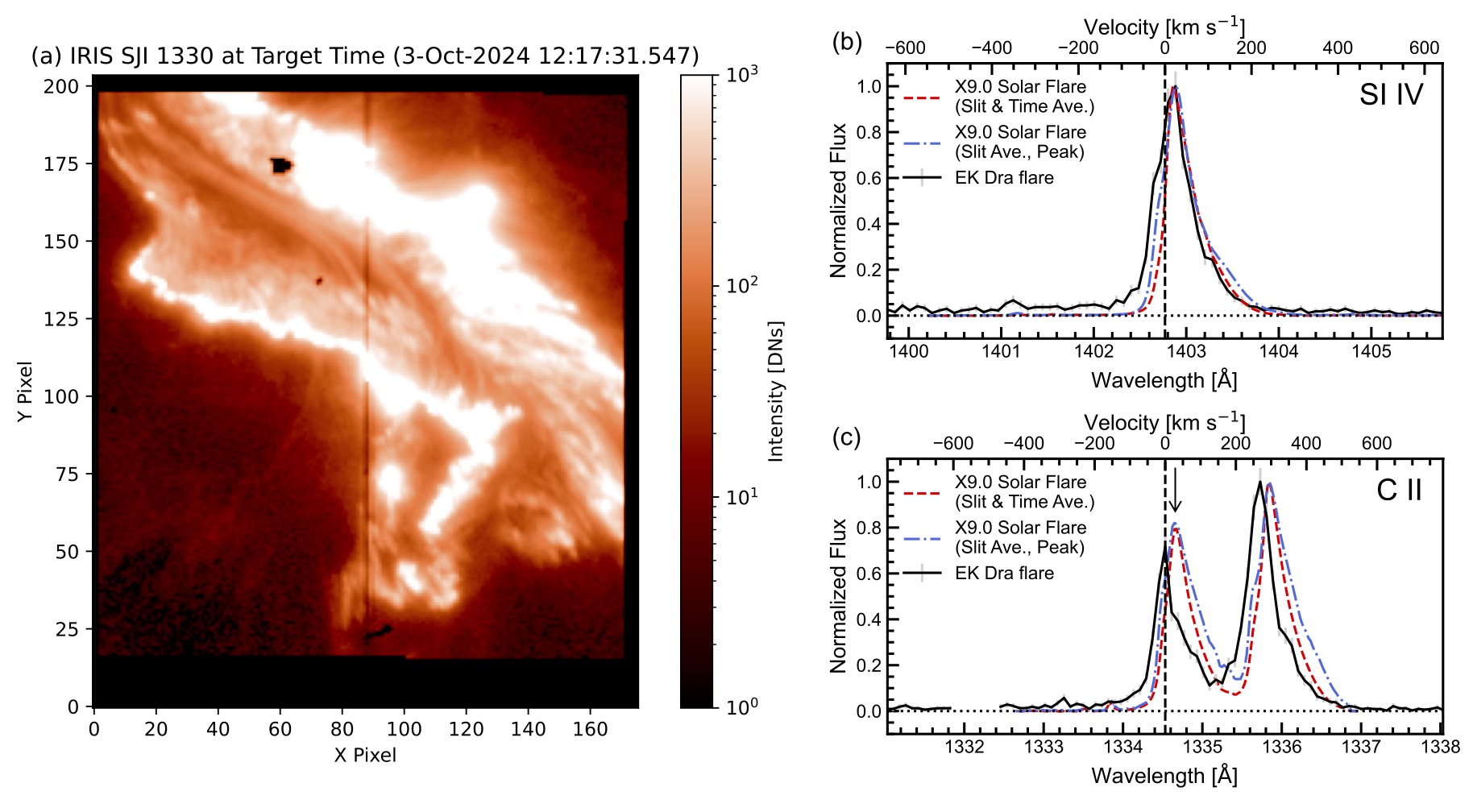}
\caption{Comparison between the stellar flare and an X9.0-class solar flare observed by IRIS on 3 October 2024 in the Si IV and C II lines.
(a) IRIS slit-jaw image taken at 12:17:31.547 near the flare peak without saturation in spectra. The vertical dark line at the center represents the spectrograph slit, which intersects both flare ribbons.
(b,c) Comparison among slit-averaged peak flare spectrum (blue), slit- and time-averaged flare spectrum (red), and the FUV flare spectrum of EK Dra (black).
\Add{Note that in panel (c), the red wing of C II 1334.53 {\AA} in the HST stellar spectra is strongly absorbed by the interstellar medium at 1334.62 {\AA} \citep{2015AJ....150....7A}, as indicated by the black arrow.}
}
\label{fig:iris}
\end{center}
\end{figure}


\begin{figure}
\begin{center}
\includegraphics[width=1.0\linewidth,clip]{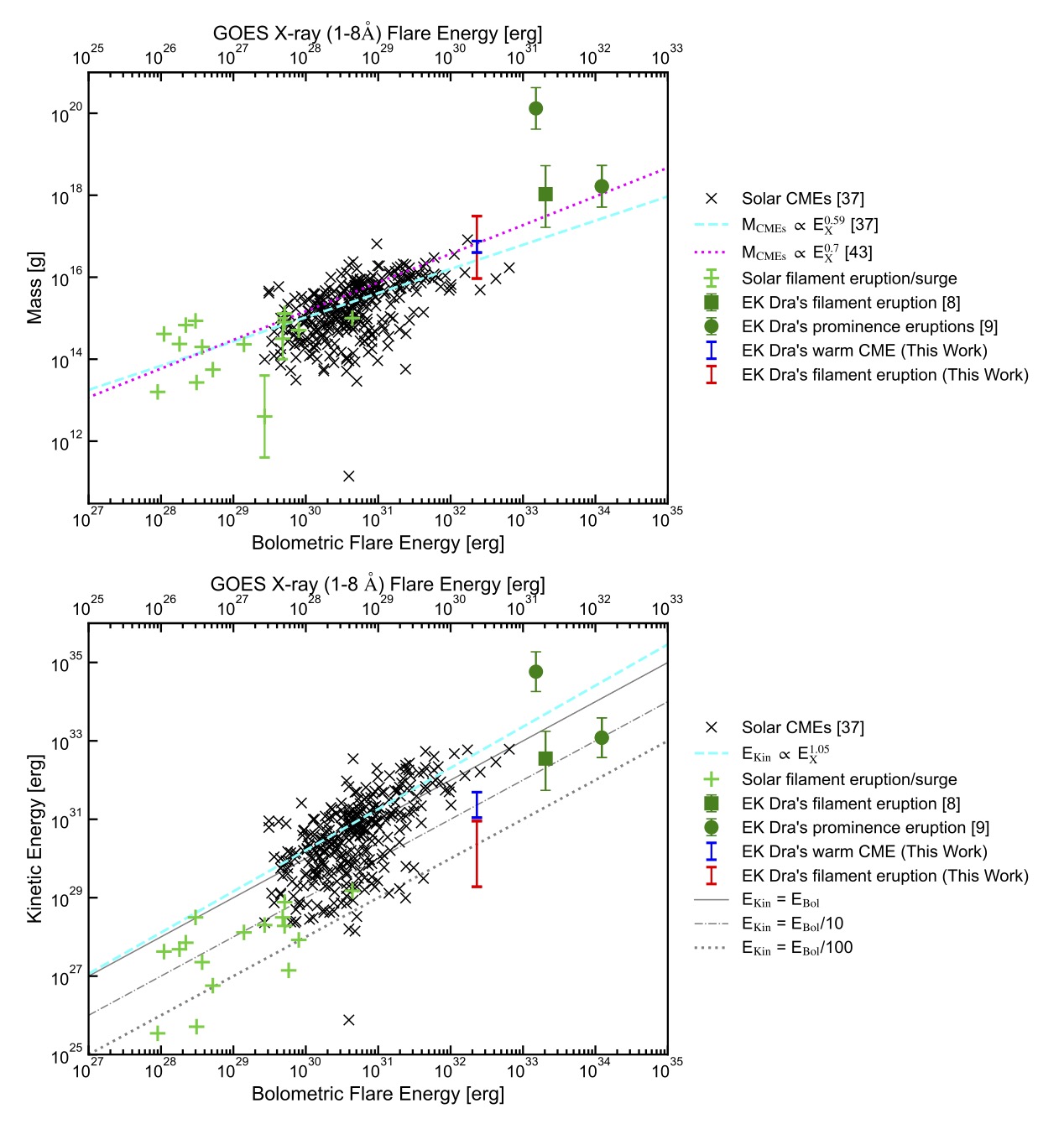}
\caption{Mass and kinetic energy of solar and stellar filament or prominence eruptions and CMEs as a function of bolometric white-light flare (WLF) energy or GOES X-ray energy. 
The red and blue symbols represent the masses of the flare event derived in this study from H$\alpha$ and FUV, respectively. The error range for H$\alpha$ originates from model parameters, whereas that for FUV arises from the Poisson error of the observed flux (see Methods).
The green points represent the stellar prominence and filament eruptions on EK Dra observed in previous study  \citep{2022NatAs...6..241N,2024ApJ...961...23N}. 
The black crosses show solar CME data \cite{2009IAUS..257..233Y,2013ApJ...764..170D}, while the green plus symbols indicate solar prominence and filament eruptions \cite{2022NatAs...6..241N,2024ApJ...961...23N}.
The cyan dashed and magenta dotted lines represent empirical relations for solar CMEs: $M_{\rm CME} \propto E_{\rm bol}^{0.59}$ and $E_{\rm kin} \propto E_{\rm bol}^{1.05}$ (cyan; \cite{2013ApJ...764..170D}), and $M_{\rm CME} \propto E_{\rm bol}^{0.7}$ (magenta; \cite{2012ApJ...760....9A}). 
Reference lines corresponding to $E_{\rm kin} = E_{\rm bol}$, $E_{\rm kin} = E_{\rm bol} / 10$, and $E_{\rm kin} = E_{\rm bol} / 100$ are shown in gray in the lower panel.
To enable direct comparison between solar and stellar data, we adopt the scaling relations $E_{\rm bol} = 100 \times E_{\rm GOES}$ \Add{\citep{2012ApJ...759...71E}} and $E_{\rm bol}$ [erg] = $10^{35} \times F_{\rm GOES}$ [W m$^{-2}$] \Add{\citep{2013PASJ...65...49S,2016A&A...588A.116W,1858MNRAS..19....1C}}. In addition, we assume that the bolometric energy is approximately equal to the bolometric WLF energy.}
\label{fig:mass}
\end{center}
\end{figure}


\clearpage


\newcommand{\noop}[1]{}

\end{document}